\documentclass[letter]{aa} 
\usepackage{graphicx}
\usepackage{txfonts}
\usepackage{natbib}
\bibpunct{(}{)}{;}{a}{}{,}
\begin{document}
   \title{Chemical depletion in the Large Magellanic Cloud: RV\,Tauri stars and the photospheric feedback
from their dusty discs.
\thanks{This research has made use of the SIMBAD database,
operated at CDS, Strasbourg, France.}}

   \author{
          C. Gielen\inst{1}
          \and
          H. Van Winckel\inst{1}
          \and
          M. Reyniers\inst{2}
          \and
          A. Zijlstra\inst{3}
          \and
          T. Lloyd Evans\inst{4}
          \and
          K.~D. Gordon\inst{5}
          \and
          F. Kemper\inst{3}
          \and
          R. Indebetouw\inst{6,12}
          \and
          M. Marengo\inst{7}
          \and
          M. Matsuura\inst{8,9}
          \and
          M. Meixner\inst{5}
          \and
          G.C. Sloan\inst{10}
          \and
          A.~G.~G.~M. Tielens\inst{11}
          \and
          P.~M. Woods\inst{3}
          }


   \institute{Instituut voor Sterrenkunde,
              Katholieke Universiteit Leuven, Celestijnenlaan 200D, 3001 Leuven, Belgium\\ 
	      \email{clio.gielen@ster.kuleuven.be}
              \and
              The Royal Meteorological Institute of Belgium, Department Observations, 
              Ringlaan 3, 1180 Brussels, Belgium 
              \and
              Jodrell Bank Centre for Astrophysics, Alan Turing Building, University of Manchester, 
              Oxford Road, Manchester, M13 9PL, United Kingdom
              \and
              SUPA, School of Physics and Astronomy, 
              University of St Andrews, North Haugh, St Andrews, Fife KY16 9SS, United Kingdom 
              \and
              Space Telescope Science Institute, 3700 San Martin Drive, Baltimore, MD 21218, USA 
              \and
              Department of Astronomy, University of Virginia, PO Box 3818, Charlottesville, VA 22903-0818, USA 
              \and
              Harvard-Smithsonian Center for Astrophysics, 60 Garden Street, MS 65, Cambridge, MA 02138-1516, USA 
              \and
              UCL-Institute of Origins, Department of Physics and Astronomy, University College London, Gower Street, London WC1E 6BT, United Kingdom 
              \and
              UCL-Institute of Origins, Mullard Space Science Laboratory, University College London, Holmbury St. Mary, Dorking, Surrey RH5 6NT, United Kingdom 
              \and
              Department of Astronomy, Cornell University, Ithaca, NY 14853-6801, USA 
              \and
              Leiden Observatory, J.H. Oort Building, Niels Bohrweg 2, 2333 CA Leiden, The Netherlands 
              \and
              National Radio Astronomy Observatory, 520 Edgemont Road, Charlottesville, VA 22906, USA 
	      \and
	      Astrophysics Group, Lennard-Jones Laboratories, Keele University, Staffordshire, ST5 5BG, United Kingdom 
}

   \date{Received ; accepted }

  \abstract
   {}
   {By studying the photospheric abundances of 4 RV\,Tauri stars in the LMC, we test whether the depletion pattern of refractory elements,
seen in similar Galactic sources, is also common for extragalactic sources. Since this depletion process probably only occurs
through interaction with a stable disc, we investigate the circumstellar environment of these sources. }
   {A detailed photospheric abundance study was performed using high-resolution UVES optical spectra. To study the circumstellar environment
we use photometric data to construct the spectral energy distributions of the stars, and determine
the geometry of the circumstellar environment, whereas low-resolution Spitzer-IRS infrared spectra are
used to trace its mineralogy.}
   {Our results show that, also in the LMC, the photospheres of RV\,Tauri stars are commonly affected by the depletion process,
although it can differ significantly in strength from source to source. From our
detailed disc modelling and mineralogy study, we find that this process, as in the Galaxy, appears closely related to the presence of a stable Keplerian disc.
The newly studied extragalactic objects have similar observational characteristics as Galactic post-AGB binaries surrounded by a dusty disc,
and are therefore also believed to be part of a binary system.
One source shows a very small infrared excess, atypical for a disc source, but still has evidence for depletion. We speculate this could point to the
presence of a very evolved disc, similar to debris discs seen around young stellar objects.}
   {}
   \titlerunning{Chemical depletion in the Large Magellanic Cloud}
   \maketitle
%

\section{Introduction}
\label{introduction}

With a distance of $\sim$\,50\,kpc \citep{feast99}, the Large
Magellanic Cloud (LMC) is one of the best galaxies to study stellar
evolution. The known distance allows us to
calculate the luminosity for stellar sources and it is the ideal
laboratory to study physical and chemical processes in an environment
with a sub-solar metallicity of $Z\sim 0.3-0.5$\,Z$_\odot$
\citep{westerlund97}. Moreover, it is close enough to allow detailed
studies of individual sources using large-aperture ground based telescopes.

Here we focus on a sample RV\,Tauri stars in the LMC.
RV\,Tauri stars are pulsating evolved stars with
a characteristic light curve showing alternating deep and shallow
minima. They are located in the high luminosity end of the Population
II Cepheid instability strip \citep{wallerstein02}. 
The post-AGB status of RV\,Tauri stars was a long standing debate, but
the detection of circumstellar dust around many objects
\citep{jura86}, and the detection of extragalactic pulsators
and their large derived absolute luminosities, were in line with the
expected evolutionary tracks of post-AGB stars. The first
extragalactic RV\,Tauri stars in the LMC were discovered by
the MACHO experiment \citep{alcock98}.

\citet{reyniers07a} and \citet{reyniers07b} performed a chemical study
on two LMC  RV\,Tauri
stars, selected from those reported by \citet{alcock98}, and found
that, like in the Galaxy \citep{vanwinckel03}, post-AGB stars are
chemically much more diverse than previously anticipated. One of the
stars, MACHO\,47.2496.8, proved to be strongly enhanced in s-process
elements, in combination with a very high carbon abundance
(C/O\,$>$\,2 and $^{12}$C/$^{13}$C\,$\approx$\,200). The metallicity
of [Fe/H]\,$=$\,$-1.4$ is surprisingly low for a field LMC star. The s-process
enrichment is large: the light s-process elements of the Sr-peak are enhanced
by 1.2 dex compared to iron ([ls/Fe]\,$=$\,$+1.2$), while for the heavy
s-process (Ba-peak) elements, an even stronger enrichment is measured:
[hs/Fe]\,$=$\,$+2.1$. Lead was not found to be strongly enhanced. The patterns
can only be understood assuming a very low efficiency of the $^{13}$C
pocket \citep{bonacic07} which
is created during the dredge-up phenomenon and the associated partial
mixing of protons into the intershell. It was the first detailed study
of the s-process of a post-AGB star in an external galaxy.

Another object, MACHO\,82.8405.15, turned out to be chemically altered
by the depletion phenomenon \citep{reyniers07a}. Depletion of refractory
elements in the photosphere is a chemical process in which chemical elements with a high dust
condensation temperature are systematically underabundant \citep[e.g.][]{maas05,giridhar05}.
The special photospheric abundance patterns are the result of gas-dust
separation in the circumstellar environment, followed by re-accretion
of only the gas, which is poor in refractory elements. The photospheres become deficient in
refractories (as Fe, Ca and the s-process elements), while the
non-refractories are not affected.
The best abundance tracers of the depletion phenomenon are the Zn/Fe and S/Ti ratios
because the elements involved in either ratio have a similar nucleosynthetic formation
channel, but have very different condensation
temperatures. Intrinsically Fe-poor objects have [Zn/Fe] and [S/Ti]
close to solar, which is not the case for depleted objects. 
With [Fe/H]\,$=$\,$-$2.6, in combination with [Zn/Fe]\,$=$\,$+$2.3 and
[S/Ti]\,$=$\,$+$2.5, in MACHO\,82.8405.15, there is no doubt
that the depletion affected the photosphere of this LMC star very
strongly. The very low abundances, as well as the clear correlation
with condensation temperature, are shown in Figure~\ref{depletie}, which is a
reproduction of the figure in \citet{reyniers07a}.

Photospheric depletion is surprisingly common in Galactic post-AGB
stars \citep[e.g.][and references therein]{giridhar05,maas05}. In
almost all depleted post-AGB objects, there is observational evidence
that a stable circumbinary disc is present
\citep{deruyter06, vanwinckel07}. The discs are very compact
\citep[e.g.][]{deroo06, deroo07c} and very likely only found around
binary post-AGB stars \citep[e.g.][and references therein]{vanwinckel09}.

Also in their infrared spectra, the RV\,Tauri stars show unique
spectral features. One of the best studied Galactic RV\,Tauri stars is AC\,Her \citep{molster99}.
Infrared studies with ISO have shown very strong crystalline silicate bands from 10--50\,$\mu$m.
Recent studies with Spitzer show that this strong crystallinity is
commonly observed, also in other post-AGB binary sources \citep{gielen08}.

\begin{figure}
\centering
\resizebox{10cm}{!}{\includegraphics{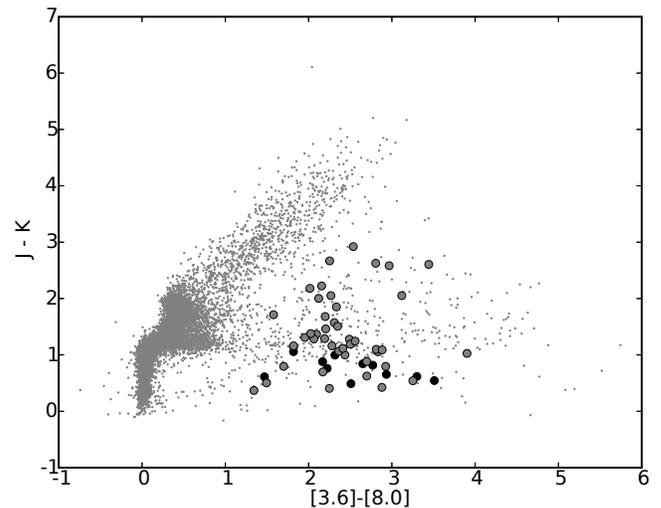}}
\caption{Colour-colour diagram indicating the Galactic post-AGB sources
  with discs (grey circles) and extragalactic RV\,Tauri stars as presented by \citet{alcock98} (black circles).
The grey dots represent the LMC objects as found by the SAGE-LMC survey \citep{meixner06}.}
\label{kleurkleur}
\end{figure}

Thanks to the efficient infrared detectors of the Spitzer satellite,
very sensitive infrared observations allow us to probe for circumstellar
dust, even around individual objects in external galaxies.
The SAGE (Surveying the Agents of a Galaxy's Evolution)
Spitzer LMC survey \citep{meixner06} mapped the LMC, using all
photometric bands of the Spitzer IRAC (InfraRed Array Camera,
\citealp{fazio04}) and MIPS (Multiband Imaging Photometer for Spitzer,
\citealp{rieke04}) instruments. This survey resulted in the detection
of over 4 million sources. Thanks to the release of this database, we
found that the LMC RV\,Tauri stars as discovered with the MACHO
experiment in the visible, indeed have infrared excesses with very
similar SED shapes as many
Galactic post-AGB binaries (see Fig.~\ref{kleurkleur}).
Since only those LMC sources with strong enough optical fluxes were selected,
some observational bias exists to stars where we see the disc more face-on,
since edge-on disc would obscure the star too much.

\begin{table*}
\caption{List of stellar parameters for our sample sources.}
\label{stelpar}
\vspace{0.cm}
\hspace{-.5cm}
\begin{tabular}{lllcrrccccc}
\hline \hline
Name & other name & $\alpha$ (J2000) & $\delta$ (J2000)  & $T_{\rm eff}$ & $\log g$ & [Fe/H] &  $E(B-V)_{tot}$ & $L_{\rm IR}/L_*$ & $L_*$     & $P$\\ 
     & &(h m s)          & ($^\circ$ ' '')   & (K)           & (cgs)    &        &                 & (\%)             & L$_\odot$ & (days)\\
\hline
MACHO\,79.5501.13 & J051418.1-691235 & 05 14 18.1 & -69 12 34.9 & 5750 & 0.5 & -2.0 & $0.14\pm0.01$ & $60\pm3$ & $5000\pm500$ & 48.5\\ 
                  & HV\,915   &&&&&&&&&\\ 
MACHO\,81.8520.15 & J053254.5-693513 & 05 32 54.5 & -69 35 13.2 & 6250 & 1.0 & -1.5 & $0.27\pm0.03$ & $0\pm1$  & $4200\pm500$ & 42.1\\
MACHO\,81.9728.14 & J054000.5-694214 & 05 40 00.5 & -69 42 14.6 & 5750 & 1.5 & -1.0 & $0.05\pm0.02$ & $53\pm3$ & $4200\pm500$ & 47.1\\
MACHO\,82.8405.15 & J053150.9-691146 & 05 31 51.0 & -69 11 46.4 & 6000 & 0.5 & -2.5 & $0.05\pm0.01$ & $84\pm3$ & $4000\pm500$ & 46.5\\
\hline
\end{tabular}
\begin{footnotesize}
\begin{flushleft}
Note: The name, equatorial coordinates $\alpha$ and $\delta$ (J2000), effective
temperature $T_{\rm eff}$, surface gravity $\log g$ and metallicity [Fe/H] of our sample stars.
Also given are the total reddening $E(B-V)_{tot}$, the energy ratio $L_{\rm IR}/L_*$, the calculated luminosity (computed by integrating
the dereddened photosphere),
assuming a distance of $d=50$\,kpc, and the period $P$ as given in \citep{alcock98}.
\end{flushleft}
\end{footnotesize}
\end{table*}

\begin{table}
\caption{\label{logUves}Log of the UVES observations and the obtained
  final S/N at a given spectral band. }
\centering
\vspace{0.5cm}
\hspace{0.cm}
\begin{tabular}{lrrrr} 
\hline\hline\rule[0mm]{0mm}{3mm}
Star               & Exp. Time & Wavelength       & S/N    \\
                   &   (sec.)    &  (nm)              &         \\  \hline
MACHO\,79.5501.13  & 3600      &  375.8 -- 498.3  & 70      \\ 
                   &           &  670.5 -- 1008.4 & 100      \\
MACHO\,81.8520.15  & 3600      &  478.0 -- 680.8  & 70    \\
MACHO\,81.9728.14  & 3600      &  478.0 -- 680.8  & 65 \\ \hline
\end{tabular}
\begin{footnotesize}
\begin{flushleft}
Note: The S/N is measured in the middle
  of the spectral window covered.
\end{flushleft}
\end{footnotesize}
\end{table}

In this contribution we focus on the MACHO RV\,Tauri objects in the
LMC as detected by the MACHO experiment \citep{alcock98}, and we
connect the chemical studies based on high-resolution optical spectroscopy to
the SED energetics and the infrared spectra obtained by Spitzer.
Prior to this study, only 1 depleted post-AGB star in the LMC was known. Here we analyse
the abundances of 4 more similar sources, testing if depletion is also
a common process in the RV\,Tauri stars of the LMC.
As the distance to the LMC is known, we are able to discuss the evolutionary 
status of these extragalactic sources using their accurate position in
the H-R diagram, which sofar has been impossible for Galactic post-AGB stars.

This research was possible thanks to the SAGE-Spec international
program (\textit{http://sage.stsci.edu/}). The goal of this large spectroscopic
infrared program is to complement the wealth of photometric data from the SAGE
photometric survey, with an extensive spectroscopic follow-up programme
using the infrared IRS spectrograph aboard of Spitzer (Kemper et al., 2009, submitted). The main goal
of the survey is to determine the composition, origin and evolution,
and observational characteristics of interstellar and circumstellar dust and its role in
the LMC. A total of about 200 stars, in all stages of stellar
evolution, HII and diffuse regions were observed in Spitzer-IRS	and MIPS-SED mode.

\section{Observations and data reduction}

\subsection{Optical high-resolution programme} 

The optical high-spectral-resolution data were obtained with the UVES
\citep{dekker00} spectrograph mounted on the Nasmith focus of UT2 of 
the VLT, in February 2005 in visitor mode under mediocre sky conditions. 

To cover a wide spectral domain, two wavelength settings of the 
spectrograph were used: one with the dichroic, to use both arms of the
spectrograph simultaneously (Dic2 860+437 nm setting) and one centered
on 580\,nm using only the red arm. The detailed log of the
observations is given in Table~\ref{logUves}. The run was plagued
with bad weather conditions, so for none of the three stars discussed
here, we could obtain the full optical spectral window.

\begin{figure*}
\centering
\rotatebox{270}{\resizebox{10cm}{!}{\includegraphics{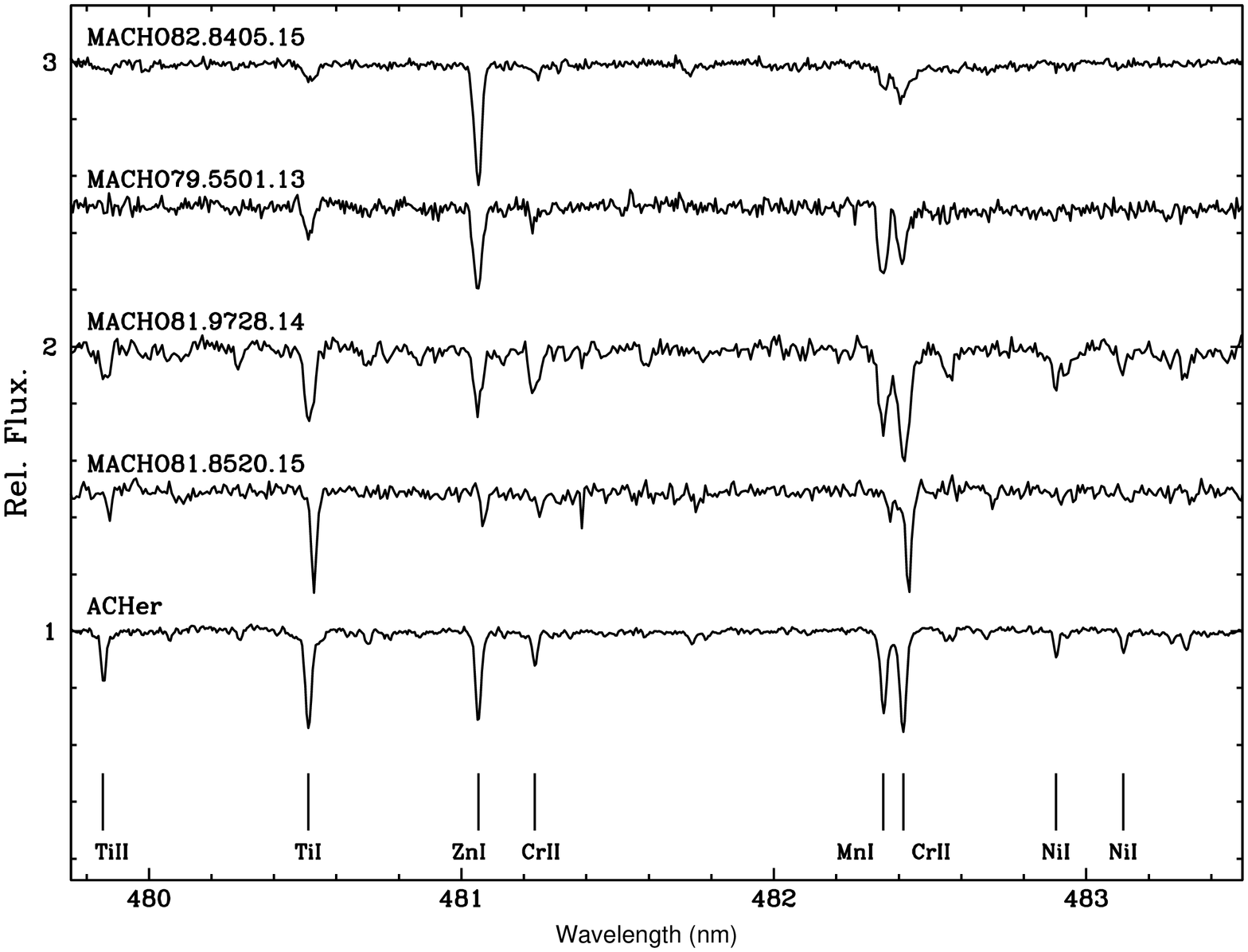}}}
\caption{Sample UVES spectra of the MACHO LMC stars. The spectra were
  normalised and offset for illustrative purposes. AC\,Her is a
  Galactic source which is a spectral analogue of the LMC sources.}
\label{lmcspec1}
\rotatebox{270}{\resizebox{10cm}{!}{\includegraphics{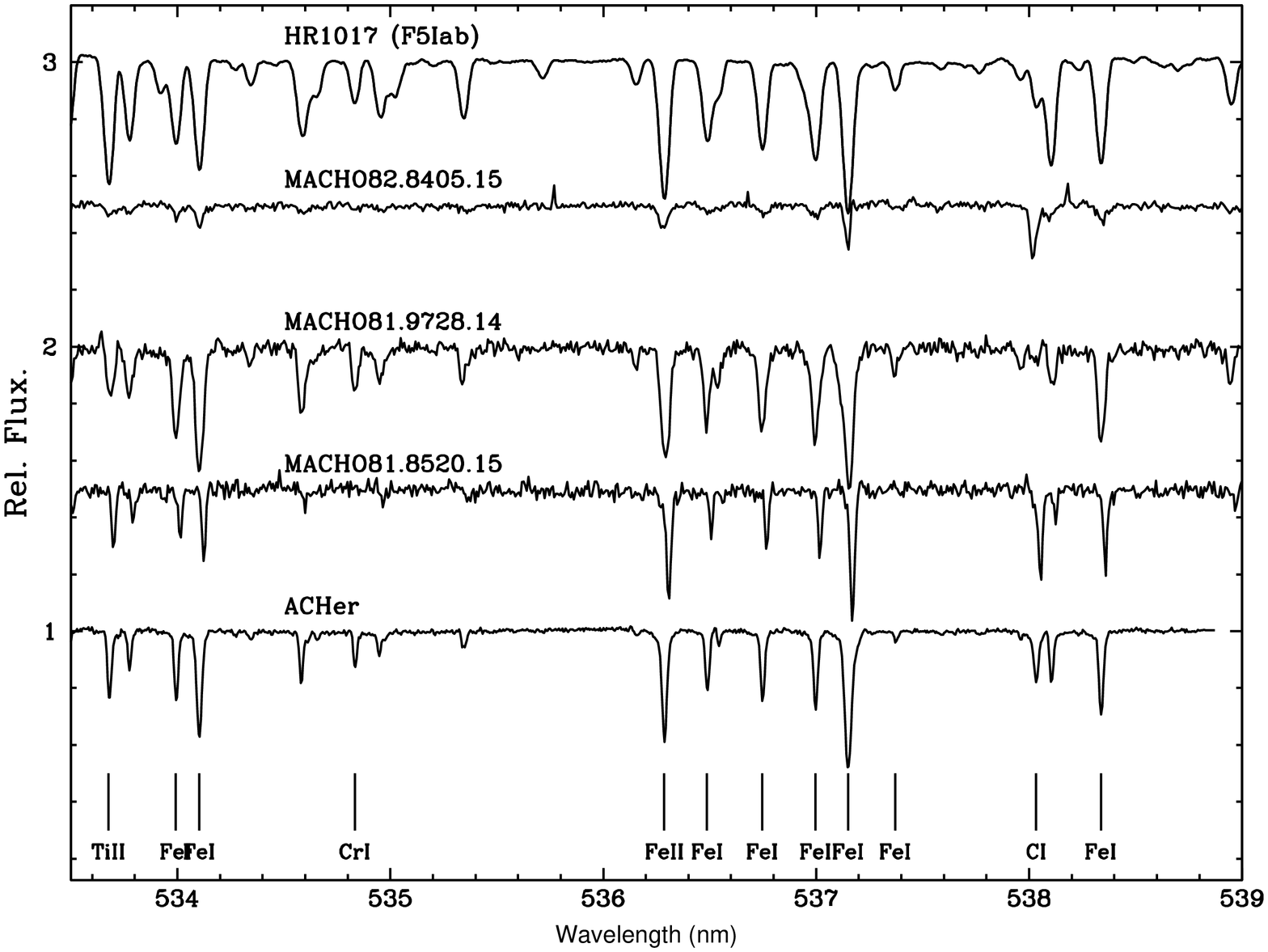}}}
\caption{Sample UVES spectra of the MACHO LMC stars. The spectra were
  normalised and offset for illustrative purposes. AC\,Her is a
  Galactic source which is a spectral analogue of the LMC sources. The
upper spectrum is from HR1017, a F5Iab star with solar composition.}
\label{lmcspec2}
\end{figure*}

The data were reduced using the dedicated UVES
context in the MIDAS software package and include the standard
reduction steps of echelle data reduction. Spectral normalisation was
performed by fitting polynomial functions through interactively
determined continuum windows. For the lower S/N data,
optimal extraction was used. Sample spectra are shown in
Figures~\ref{lmcspec1} and \ref{lmcspec2}.

\subsection{IRS-spectroscopy}

\begin{figure}
\centering
\resizebox{8cm}{!}{\includegraphics{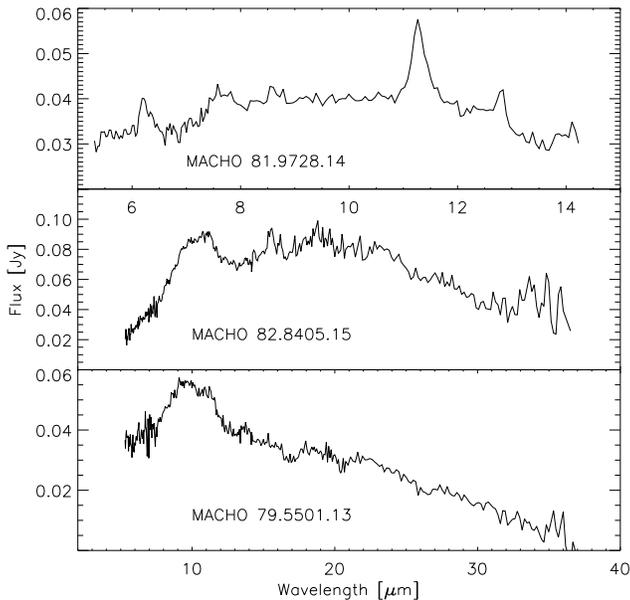}}
\caption{Spitzer-IRS spectra of MACHO\,79.5501.13, MACHO\,81.9728.14 and MACHO\,82.8405.15.}
\label{plot_machos}
\end{figure}

In  the follow-up programme SAGE-Spec to the SAGE LMC survey
several post-AGB sources, showing RV\,Tauri-like behaviour, were
observed with the IRS instrument aboard the Spitzer Space Telescope
\citep{houck04,werner04}. IRS was used in low-resolution SL and LL spectroscopic
staring modes. SL ($\lambda$=5.3-14.5\,$\mu$m) and LL
($\lambda$=14-37\,$\mu$m) spectra have a resolving power of
R=$\lambda/\bigtriangleup\lambda \sim$ 100. Exposure times were
chosen to achieve a S/N ratio of around 60 for the SL modes, LL modes
have a S/N of 30. For a detailed description of the observations we refer to
Kemper et al. (in prep.). Spitzer-IRS spectra for three of our sample stars are
shown in Figure~\ref{plot_machos}.

The extraction was done from intermediate droop data products, version
S17.0.4 up to S18.0.2, as delivered by the Spitzer Science Center. SMART reduction
package tools \citep{higdon04}, and reduction tools developed by the
FEPS (the Spitzer legacy program ``Formation and evolution of
planetary systems'') team, were used for further extraction. For a detailed description of the different
reduction steps we refer to \citet{hines05}. The spectra are
background-corrected and bad/hot-pixel-corrected and a fixed-width
aperture is used for the extraction. The spectra are then combined
and order matched. For the defringing of the spectra the IRSFRINGE \citep{lahuis03}
package was used. Calibration spectral response functions were
calculated, derived from standard stars and corresponding stellar
models.

In this contribution we limit the IRS spectral analyses to those
objects for which we also have UVES data.

\onecolumn
\begin{figure}
\vspace{0cm}
\includegraphics[width=9cm,height=6cm]{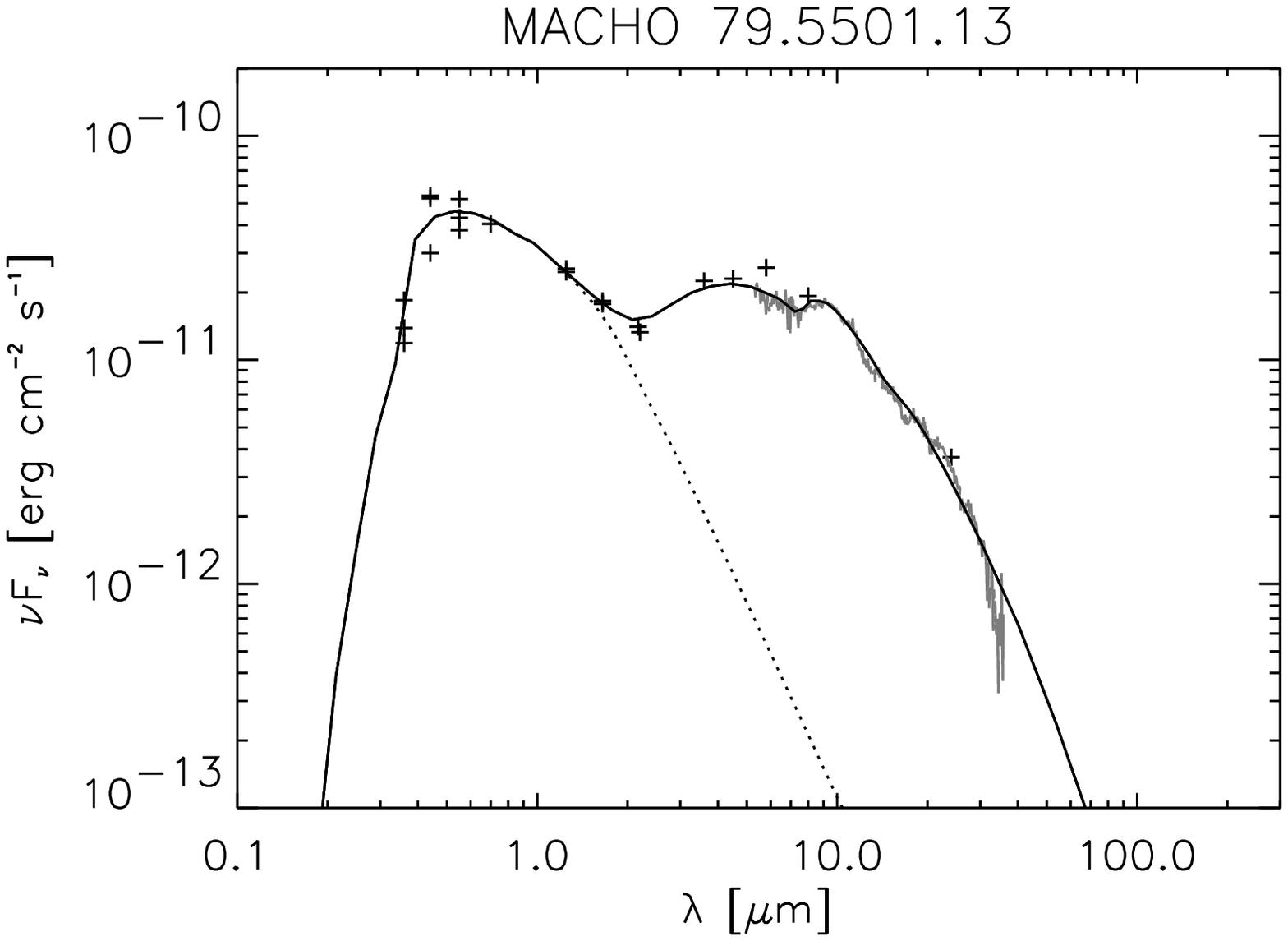}
\includegraphics[width=9cm,height=6cm]{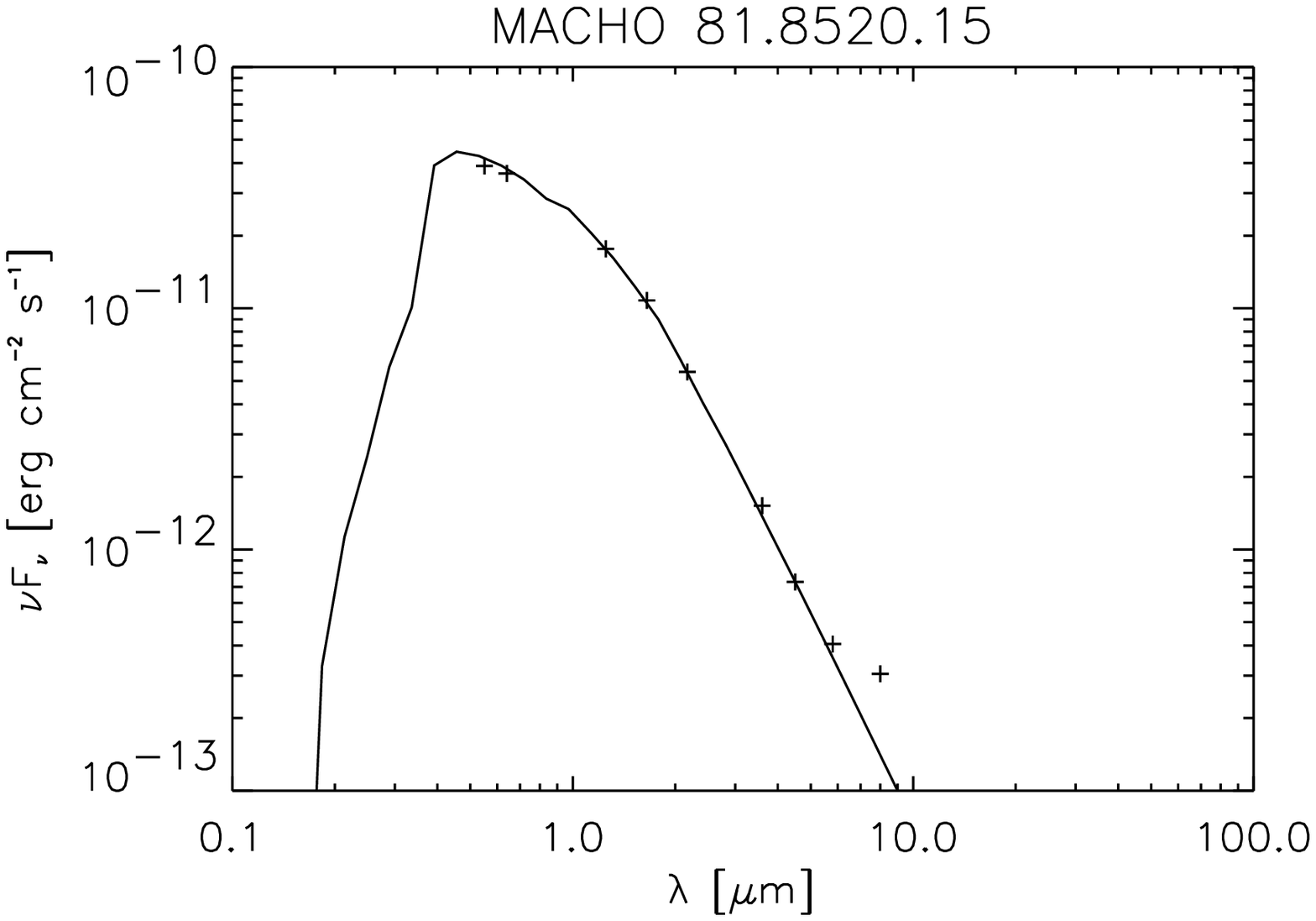}
\end{figure}
\begin{figure}
\vspace{-0.65cm}
\hspace{0.8cm}
\includegraphics[width=8.2cm,height=5cm]{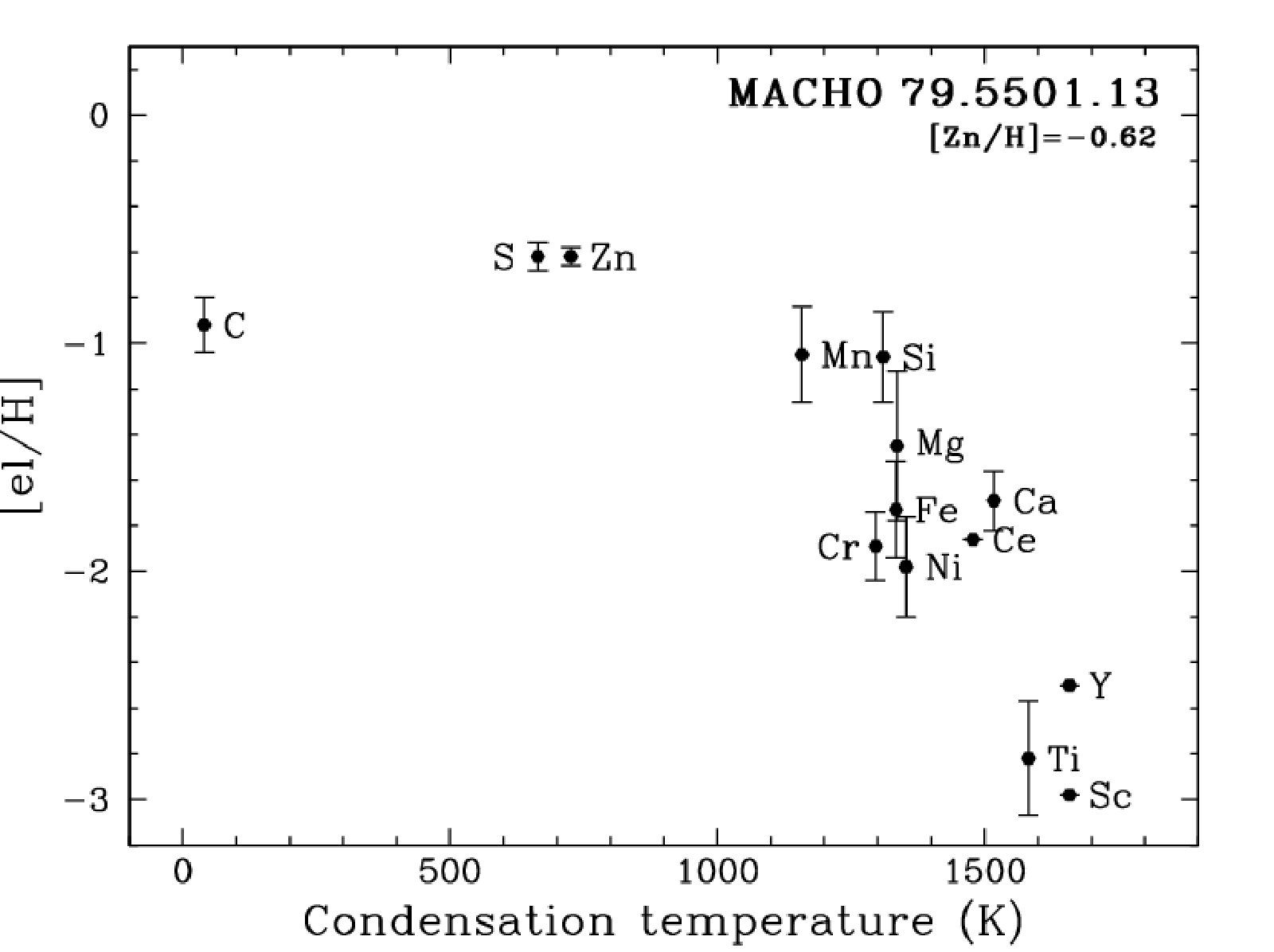}
\hspace{0.7cm}
\includegraphics[width=8.2cm,height=5cm]{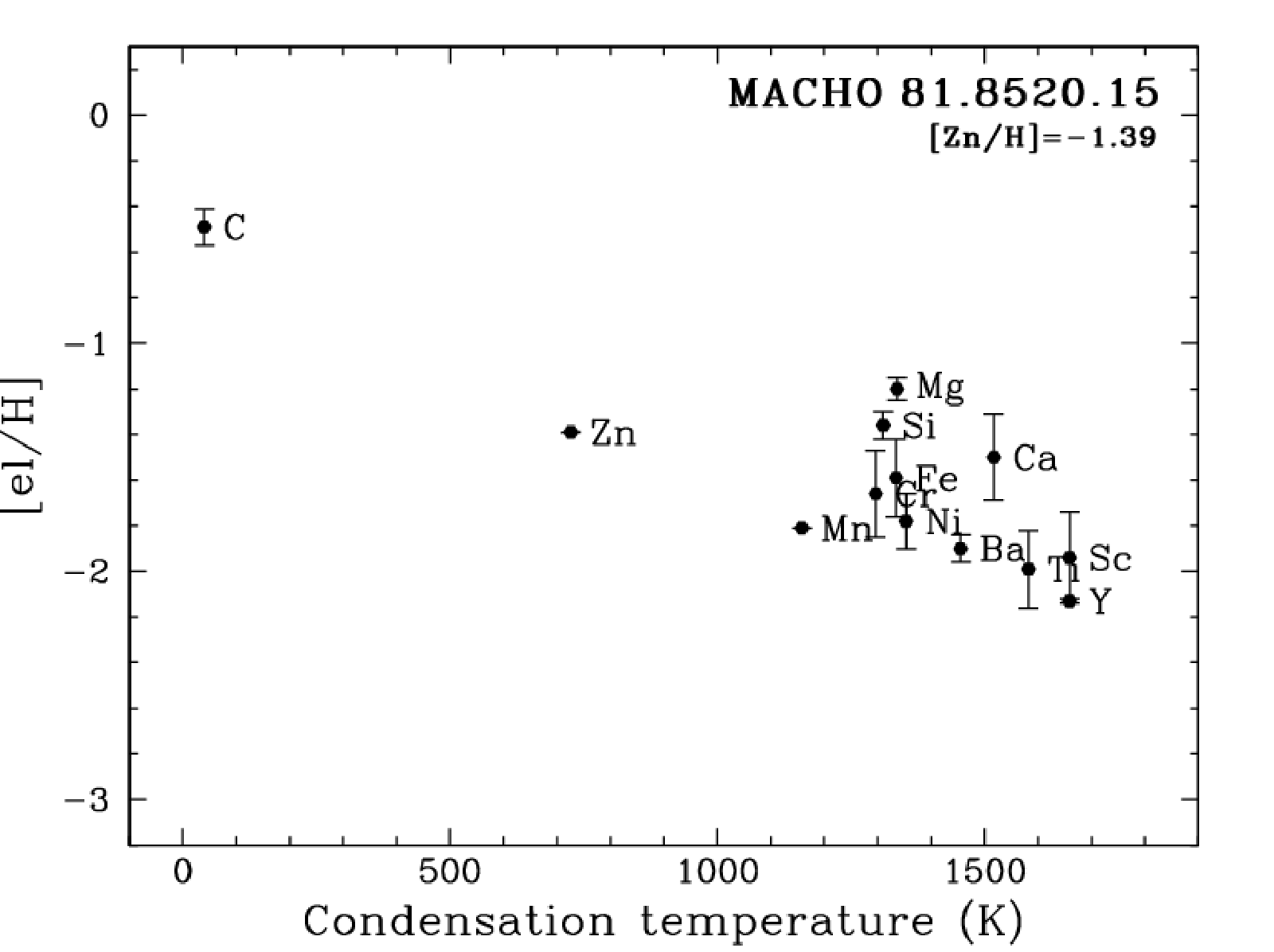}
\end{figure}
\begin{figure}
\vspace{0cm}
\includegraphics[width=9cm,height=6cm]{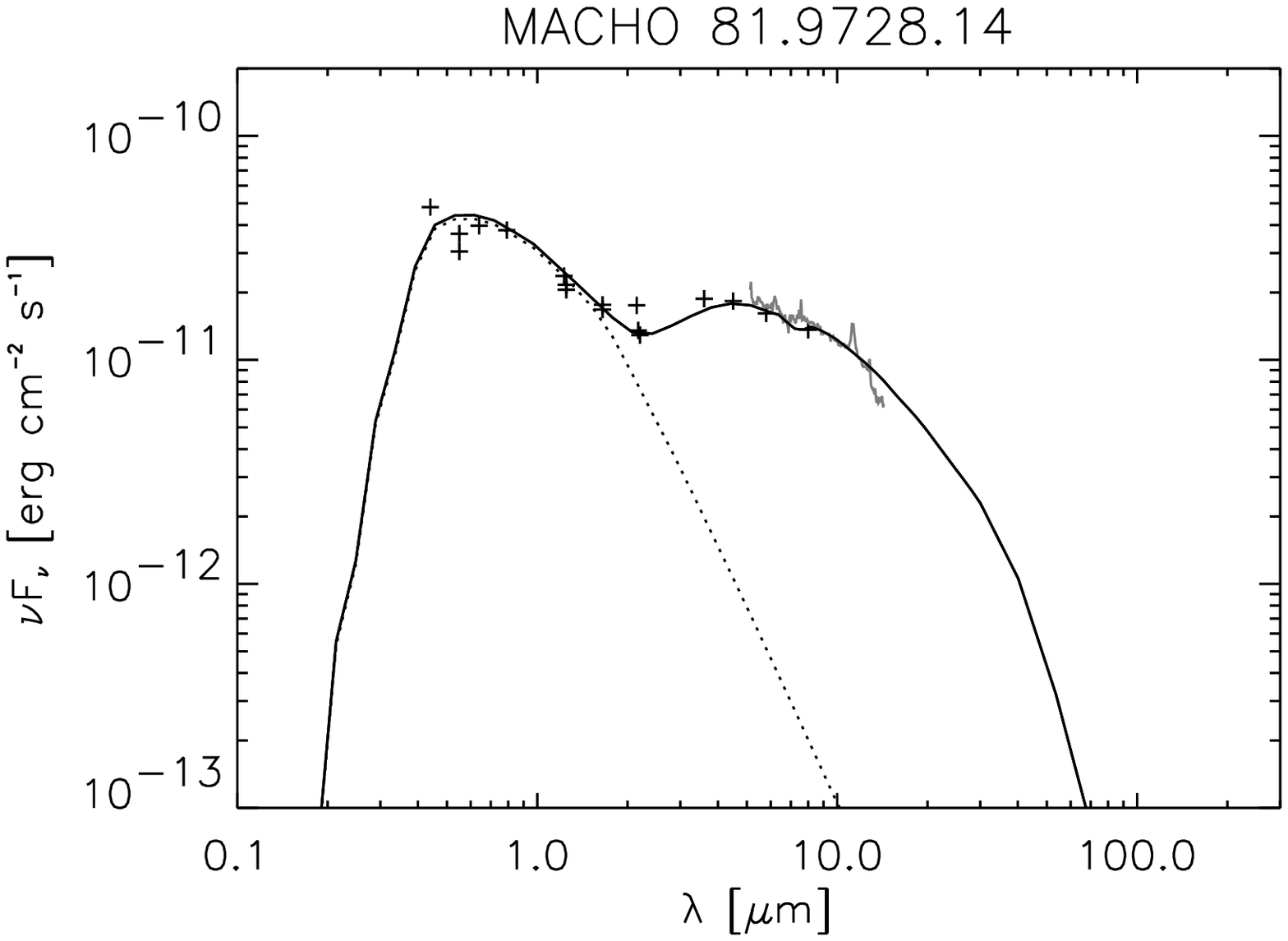}
\includegraphics[width=9cm,height=6cm]{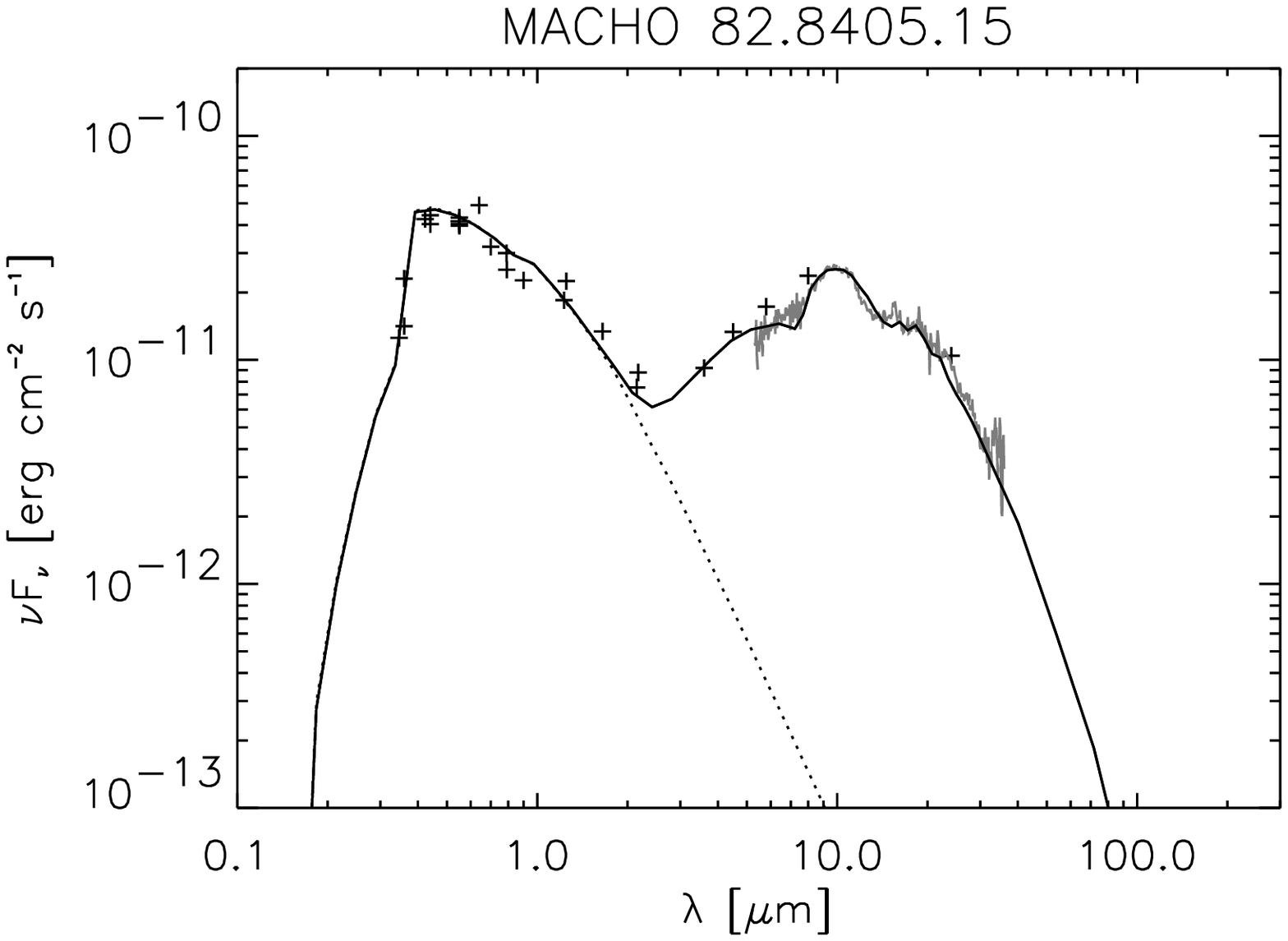}
\end{figure}
\begin{figure}
\vspace{-0.65cm}
\hspace{0.8cm}
\includegraphics[width=8.2cm,height=5cm]{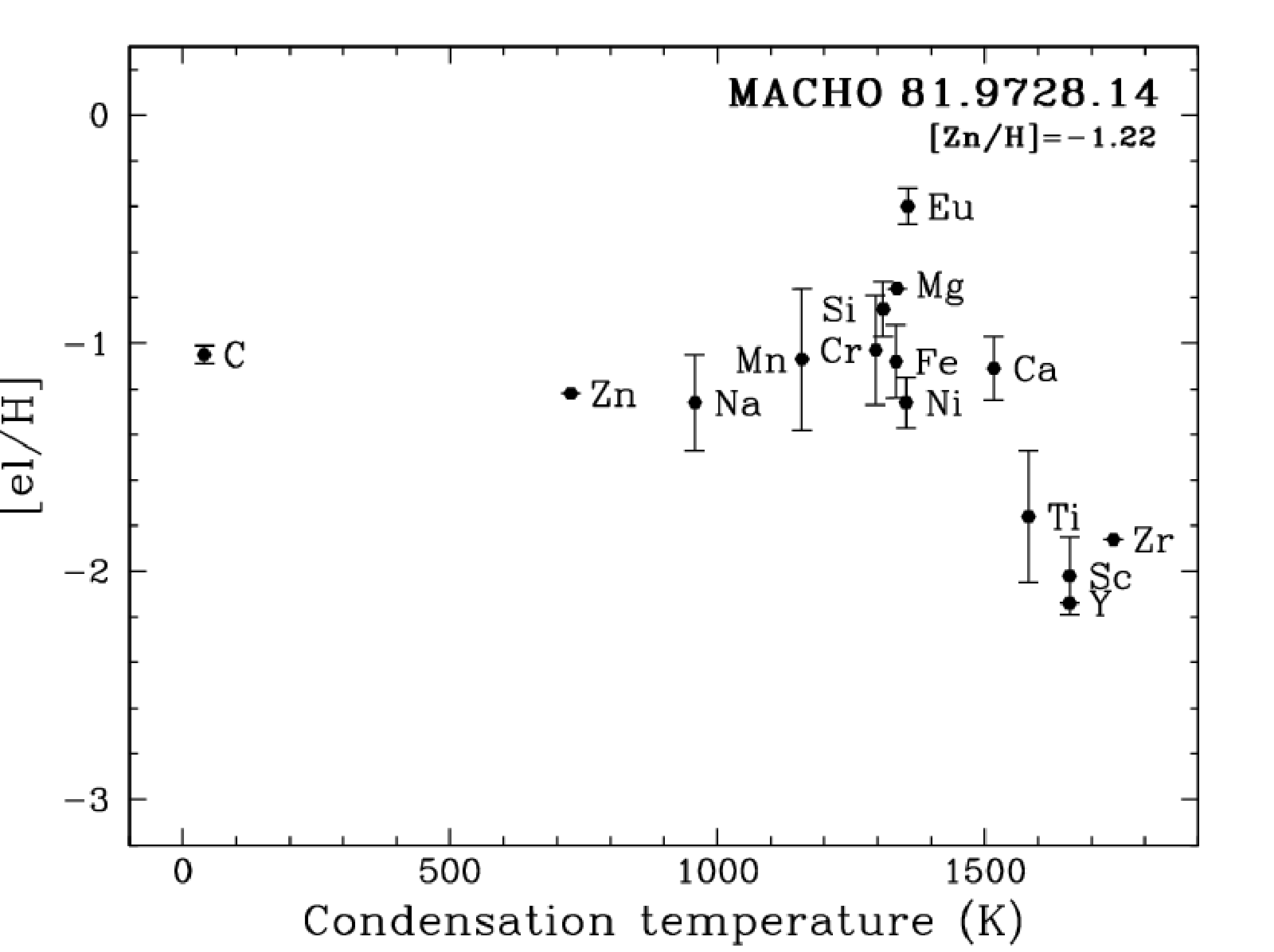}
\hspace{0.7cm}
\includegraphics[width=8.2cm,height=5cm]{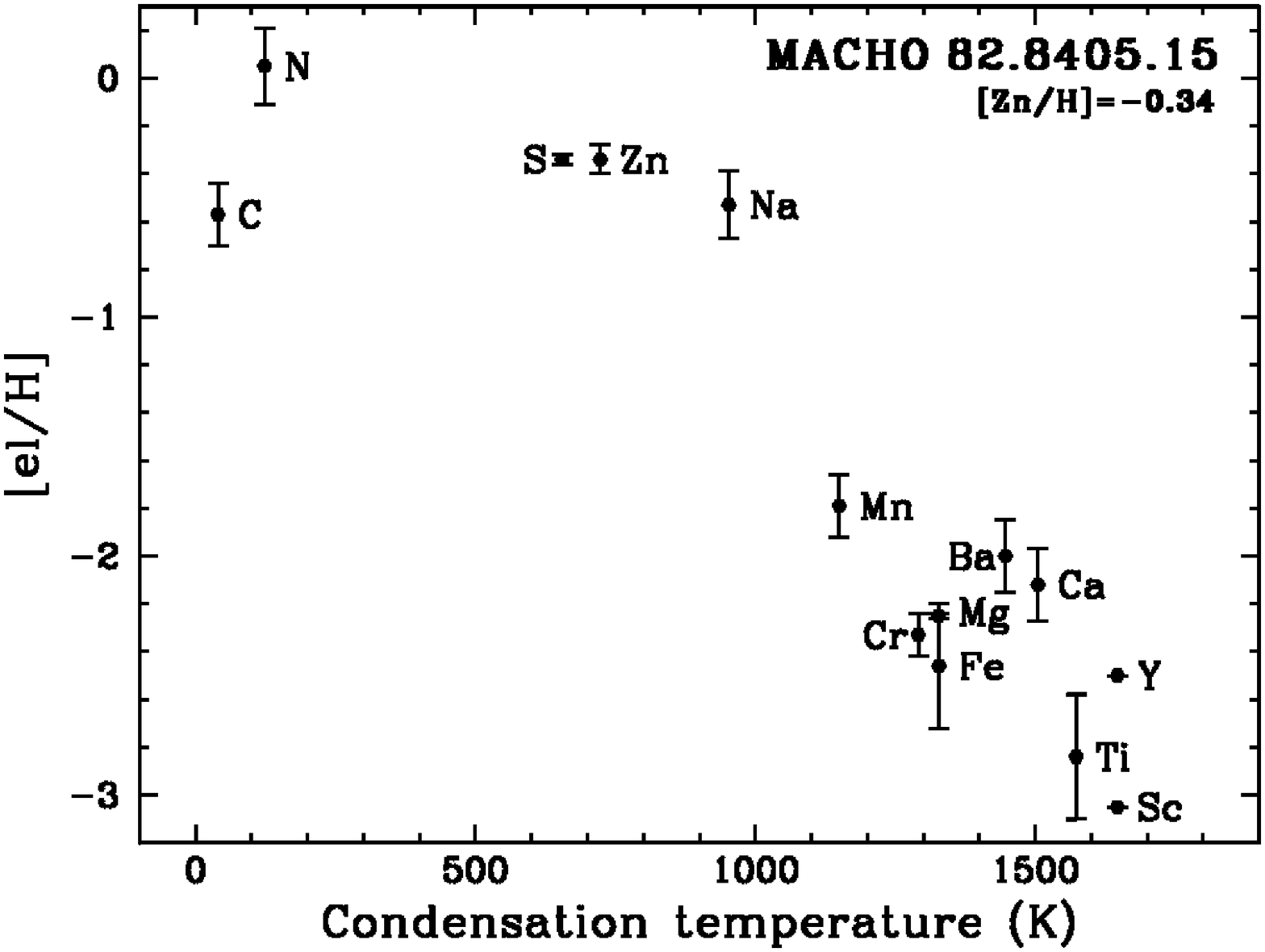}
\caption{First and third row: The spectral energy distributions of our sample stars. Crosses represent photometric data, and the observed Spitzer spectrum is overplotted in gray. The solid line give the homogeneous disc model as discussed in Section~\ref{sedmodel}. The dotted line gives the Kurucz stellar model.
For MACHO\,81.8520.15 we only show the photometric data (crosses) and the appropriate Kurucz model (solid line).
Second and fourth row: Below the corresponding SED we also show the abundance pattern [el/H] of the stars as a function
of the condensation temperature.}
\label{depletie}
\end{figure}
\twocolumn

\section{Photospheric abundance results}

One of the features of the spectra displayed in
Figures~\ref{lmcspec1} and \ref{lmcspec2} 
is the general weakness of metallic lines. The LMC RV\,Tauri stars are proven
to be spectral analogues of the Galactic RV\,Tauri star AC\,Her, also
shown in the figures. This Galactic RV\,Tauri star is considered to be one of the prototypes
of its pulsation class, but it is also known to be a strongly depleted object
\citep{giridhar89,vanwinckel98}.

Our analysis of the final product spectra is similar to those
used in \citet{vanwinckel00} and \citet{reyniers07b}. We performed a
line identification using the solar line lists of \citet{thevenin89,thevenin90}. 
Weak lines, on the linear part of the curve-of-growth are the best
tracers of the chemical conditions of the photospheres, as they are not
saturated and are formed deeper in the photosphere, where possible non-LTE
effects are known to be weaker. 
For the chemical analysis, we only used lines with
strength below 150\,m\AA{} and accurate oscillator strengths
\citep{vanwinckel00}. 

As the stars seem chemically peculiar, we
performed a relative abundance analysis using the Galactic RV\,Tauri
star AC\,Her. The quantified abundance analysis was performed using the latest
Kurucz model atmospheres of the appropriate metallicity
(\textit{http://kurucz.harvard.edu/}), and the LTE abundance analysis
  program MOOG (April 2002 version) written by Prof. C. Sneden (\textit{http://verdi.as.utexas.edu/moog.html}).

\subsection{MACHO\,79.5501.13}
\subsubsection{Photospheric model}
To determine the most appropriate photospheric model, we combined the
classical spectroscopic methods with Balmer line-profile fitting and
the constraints from the star being a member of the LMC. We refer to
\citet{reyniers07b} for a more detailed description of the methods.

The spectroscopic analysis assumes that the excitation of the Fe I
lines follow a Boltzman distribution and by requiring that the
abundances of lines from low and high excitation levels must
be the same, a spectroscopic temperature is deduced. The gravity is
determined by imposing that the abundance of different ions must be the
same \citep[e.g.][]{vanwinckel00}. For pulsating stars with large
amplitude pulsations and associated shocks, this method turns out to
be too sensitive to non-LTE effects. We therefore refined the
effective temperature and gravity determination by fitting the Balmer
$\delta$ and $\gamma$ line profiles with synthetic models. As the
object is known to be a member of the LMC, we assumed the distance to be 50\,kpc, and determined the absolute luminosity after
dereddening of the colours. A surface gravity
estimate is then deduced using a standard mass for
a post-AGB star of 0.6\,M$_{\odot}$. With this process, we find best values of
$T_{\rm eff}=5750$\,K, $\log g = 0.5$ and a [Fe/H] abundance of $-1.7$.
Typical errors are $\pm 250$\,K for the $T_{\rm eff}$, and 0.5 for the $\log g$.

\subsubsection{Chemical analysis}

The results of our quantitative analyses are given in Table~\ref{abundancerslts} and
displayed in Figure~\ref{depletie}. The spectral coverage of this
object is not optimal as the intermediate spectral window of UVES with
central wavelength of 580\,nm was not observed. 
The results of our chemical study clearly confirm that the object
is a spectral analogue to AC\,Her. Its Fe abundance is low
([Fe/H]\,$=$\,$-1.8$), even for the LMC, but this does not reflect the
initial abundance. A clear correlation is found when depicting the
abundance of a given element to its condensation temperature
(Fig.~\ref{depletie}): refractory elements are strongly
depleted. Zn and S may be tracing the original metallicity which
was likely to be around $-0.6$. This indicates that the object was originally slightly metal deficient
compared to the overall metallicity of the LMC.

Also in this LMC source, the depletion is found to be affecting the chemical content of the photosphere very strongly. 
For Sc, an element with one of the highest condensation temperatures, the abundance is
about 250 times lower than what we would expect from a star of this
initial metallicity ([Fe/H]$_{\rm original}$\,$=$\,$-0.6$). Also Y and Ce,
the only two s-process elements for which we found lines suitable for
abundance determination, the underabundance is a factor 80 and 20
respectively.

\subsection{MACHO\,81.8520.15}
\subsubsection{Photospheric model}

The spectrum of MACHO\,81.8520.15 is of high quality which is reflected in
the number of lines used in our chemical analyses and in a
smaller line-to-line scatter. 

The star is found to be somewhat hotter than previous objects. This is also seen 
directly in the spectra, more specifically in the ratio of line strengths between lines of ionised versus neutral ions of the same species. 
Our best spectroscopic
estimate is a $T_{\rm eff}= 6250$\,K with a gravity of $\log g =1.0$.

\subsubsection{Chemical analysis}

Also in MACHO\,81.8520.15, the abundances are significantly lower than the mean
metallicity of the LMC. Assuming the Zn abundance reflects the initial
condition, a [Zn/H] = $-$1.4 makes MACHO\,81.8520.15 too a star of rather low
initial metallicity.

The abundance pattern shown in Figure~\ref{depletie} does show a
general trend with condensation temperature. The main difference is
that the C abundance turned out to be very high. The carbon
abundance is determined on the basis of 8 lines which show little
internal scatter, so the high C abundance is certainly significant.

The low Zn abundance can be interpreted either as the initial value,
which means that the object increased its C abundance by a factor of
10, or it is the C abundance which reflects the initial conditions. 
In this case the depletion has been efficient to such low
temperatures that even the Zn abundance was affected.

\begin{table*}
\vspace{-0.2cm}
\caption{Quantified abundance results for MACHO\,79.5501.13, MACHO\,81.9728.14
and MACHO\,81.8520.15.}
\label{abundancerslts}
\centering
\hspace{0.cm}
\begin{small}
\begin{tabular}{l|rrrr|rrrr|rrrr|rr}
 & 
\multicolumn{4}{c|}{\rule[-0mm]{0mm}{5mm}{\bf MACHO\,79.5501.13}}&
\multicolumn{4}{c|}{\rule[-0mm]{0mm}{5mm}{\bf MACHO\,81.9728.14}}&
\multicolumn{4}{c|}{\rule[-0mm]{0mm}{5mm}{\bf MACHO\,81.8520.15}}& & \\
 &
\multicolumn{4}{c|}{
$
\begin{array}{r@{\,=\,}l}
T_{\rm eff} & 5750\,{\rm K}  \\
\log g  & 0.5\ {\rm(cgs)} \\
\xi_{\rm t} & 3.0\ {\rm km\,s}^{-1} \\
\end{array}
$
}
&
\multicolumn{4}{c|}{
$
\begin{array}{r@{\,=\,}l}
T_{\rm eff} & 5500\,{\rm K}  \\
\log g  & 1.0\ {\rm(cgs)} \\
\xi_{\rm t} & 3.0\ {\rm km\,s}^{-1} \\
\end{array}
$
}
&
\multicolumn{4}{c|}{
$
\begin{array}{r@{\,=\,}l}
T_{\rm eff} & 6250\,{\rm K}  \\
\log g  & 1.0\ {\rm(cgs)} \\
\xi_{\rm t} & 3.0\ {\rm km\,s}^{-1} \\
\end{array}
$
} & &
\\

\hline
ion & N &$\log\epsilon$&$\sigma_{\rm ltl}$&[el/H]& N &$\log\epsilon$&$\sigma_{\rm ltl}$&[el/H]& N &$\log\epsilon$&$\sigma_{\rm ltl}$&[el/H]& Sun & T$_{\rm cond}$\\
\hline
C \,{\sc i  } &  6 &  7.65 & 0.12 & -0.92 &  2 & 7.52 & 0.04 & -1.05 &   8 & 8.08 & 0.08 & -0.49 & 8.57 &   40. \\
Na\,{\sc i  } &    &       &      &       &  2 & 5.07 & 0.21 & -1.26 &     &      &      &       & 6.33 &  958. \\
Mg\,{\sc i  } &  2 &  6.09 & 0.33 & -1.45 &  1 & 6.78 &      & -0.76 &   2 & 6.34 & 0.05 & -1.20 & 7.54 & 1336. \\
Si\,{\sc i  } &  2 &  6.48 & 0.20 & -1.06 &  4 & 6.69 & 0.12 & -0.85 &     &      &      &       & 7.54 & 1310. \\
Si\,{\sc ii } &    &       &      &       &    &      &      &       &   2 & 6.18 & 0.06 & -1.36 & 7.54 & 1310. \\
S \,{\sc i  } &  5 &  6.71 & 0.06 & -0.62 &    &      &      &       &     &      &      &       & 7.33 &  664. \\
Ca\,{\sc i  } &  2 &  4.67 & 0.13 & -1.69 &  9 & 5.25 & 0.14 & -1.11 &  12 & 4.86 & 0.19 & -1.50 & 6.36 & 1517. \\
Sc\,{\sc ii } &  1 &  0.19 &      & -2.98 &  2 & 1.15 & 0.17 & -2.02 &   5 & 1.23 & 0.20 & -1.94 & 3.17 & 1659. \\
Ti\,{\sc i  } &    &       &      &       &    &      &      &       &   1 & 3.13 &      & -1.89 & 5.02 & 1582. \\
Ti\,{\sc ii } &  6 &  2.20 & 0.25 & -2.82 &  4 & 3.26 & 0.29 & -1.76 &  16 & 3.03 & 0.18 & -1.99 & 5.02 & 1582. \\
Cr\,{\sc i  } &  3 &  3.85 & 0.02 & -1.82 &  3 & 4.39 & 0.30 & -1.28 &   4 & 3.91 & 0.16 & -1.76 & 5.67 & 1296. \\
Cr\,{\sc ii } & 10 &  3.76 & 0.16 & -1.91 & 11 & 4.70 & 0.18 & -0.97 &  12 & 4.05 & 0.19 & -1.62 & 5.67 & 1296. \\
Mn\,{\sc i  } &  6 &  4.34 & 0.21 & -1.05 &  2 & 4.32 & 0.31 & -1.07 &   1 & 3.58 &      & -1.81 & 5.39 & 1158. \\
Fe\,{\sc i  } & 41 &  5.85 & 0.15 & -1.66 & 73 & 6.43 & 0.16 & -1.08 & 122 & 5.92 & 0.17 & -1.59 & 7.51 & 1334. \\
Fe\,{\sc ii } & 14 &  5.56 & 0.22 & -1.95 & 14 & 6.47 & 0.15 & -1.04 &  23 & 5.95 & 0.15 & -1.56 & 7.51 & 1334. \\
Ni\,{\sc i  } &  2 &  4.27 & 0.22 & -1.98 & 10 & 4.99 & 0.11 & -1.26 &   6 & 4.47 & 0.12 & -1.78 & 6.25 & 1353. \\
Zn\,{\sc i  } &  3 &  3.98 & 0.04 & -0.62 &  1 & 3.38 &      & -1.22 &   1 & 3.21 &      & -1.39 & 4.60 &  726. \\
Y \,{\sc ii } &  1 & -0.26 &      & -2.50 &1&0.49& & -1.75 &   2 & 0.11 & 0.01 & -2.13 & 2.24 & 1659. \\
Zr\,{\sc ii } &    &       &      &       &  1 & 0.74 &      & -1.86 &     &      &      &       & 2.60 & 1741. \\
Ba\,{\sc ii } &    &       &      &       &    &      &      &       &   4 & 0.23 & 0.06 & -1.90 & 2.13 & 1455. \\
Ce\,{\sc ii } &  1 & -0.28 &      & -1.86 &    &      &      &       &     &      &      &       & 1.58 & 1478. \\
Eu\,{\sc ii } &    &       &      &       &  2 & 0.12 & 0.08 & -0.40 &     &      &      &       & 0.52 & 1356. \\
\hline
\multicolumn{15}{l}
{}
\end{tabular}
\end{small}
\begin{footnotesize}
\begin{flushleft}
Note: The explanation of the columns is as follows: N gives the
number of lines used; $\log\epsilon$ is the absolute abundance derived
$\log\epsilon$\,$=$\,$\log$(N(X)/N(H))+12; $\sigma_{\rm ltl}$ is the
line-to-line scatter; and [el/H] is the abundance relative to the Sun. For the
references of the solar abundances needed to calculate the [el/H] values: see
\cite{reyniers07a}. The dust condensation temperatures are taken from
\cite{lodders03}, and references therein. They are computed using a solar
abundance mix at a pressure of 10$^{-4}$\,atm. The abundance table of
MACHO\,82.8405.15 can be found in \cite{reyniers07b}. 
\end{flushleft}
\end{footnotesize}
\end{table*}

\subsection{MACHO\,81.9728.14}
\subsubsection{Photospheric model}

We obtained a good spectroscopic solution for the model atmosphere
parameters, showing that this star is a slightly cooler analogue of the
previous object. The only Balmer lines covered are H$_{\beta}$ and H$_{\alpha}$
and these are affected by emission cores which make them unsuitable
for model parameter determination. 

\subsubsection{Chemical analysis}

The chemical pattern displayed by this star is depicted in
Figure~\ref{depletie}. This object has the poorest quality spectrum,
and the line-to-line scatter in the abundances remains quite high.

The abundance distribution is 
different in this object, since we find the Zn abundance to be very low
([Zn/H]\,$=$\,$-1.22$). Interpreting this as the initial chemical condition, this
star is intrinsically of low metallicity, even for the LMC. The abundance pattern is flat
except for the refractory elements with the highest condensation
temperature. These refractories show abundances which are
significantly lower, down by up to $-0.8$ relative to the
[Zn/H] abundance. Therefore, we interpret these abundances as affected by
depletion. The effect has been marginal and only detectable for
elements with the highest condensation temperature.

The metallicity dependence of the depletion phenomenon has been found
in the Galactic sample as well \citep{giridhar05, maas05}, and may be
confirmed here: at low initial metallicity, the effect is strongly
reduced and often only visible for elements with the highest
condensation temperature.

\section{Infrared spectroscopy}
\label{features}

Depletion was found to be very efficient in the RV\,Tauri stars 
for which we have UVES spectra available. In the Galaxy, depletion patterns are 
abundant but only around stars where stable dusty discs are present
(with some noticeable exceptions like BD+39.4926 \citep{vanwinckel95}). We therefore 
used the infrared catalogue released by the SAGE team \citep{meixner06} to investigate the spectral 
energy distributions. We used the photospheric parameters from our UVES study and 
dereddened the data until the match between the photometry and the Kurucz photospheric 
model was optimal. The SED figures are shown in Figure~\ref{depletie}.
It is clear that also around these objects, which were selected on the basis of their 
light curve, circumstellar dust is present with specific colours
typical of discs \citep{deruyter06}. 
In the next sections we discuss our analysis of the spectrophotometric data we obtained for these objects. 

\subsection{MACHO\,81.9728.14}

For this star we unfortunately only possess spectral information in the
$5-14$\,$\mu$m region. The spectrum is unique in our sample, and is dominated by strong emission
features at $6.2-7.6-8.6-11.3-12.8$ and 14\,$\mu$m, which can be
identified as PAH emission \citep[][and references
therein]{tielens08}.

The C-C stretching and bending modes produce features with typical
central wavelengths at 6.2 and 7.7\,$\mu$m. The 8.6\,$\mu$m feature is
due to C-H in-plane bending modes and features longward of 10\,$\mu$m
can be attributed to C-H out-of-plane bending modes.

Following the classification as described in \citet{peeters02}, which is based
on the central wavelength of the main emission bands, the PAH
emission bands are characterised as being of class A. This class A is
usually linked to very processed PAHs residing in interstellar
material, directly illuminated by a star.

PAH bands are seen in the infrared spectra of several post-AGB
sources, but surprisingly almost always together with features of
silicate dust species. Such mixed chemistry objects include HR\,4049
\citep{johnson99,dominik03,antoniucci05,hinkle07}, and HD\,44179,
the central star of the `Red Rectangle' \citep{menshchikov02,cohen04}.
In these stars the PAH emission is classified as of class B. Class B
sources are mainly associated with circumstellar material. HR\,4049
and HD\,44179 are both binary post-AGB stars surrounded by an O-rich
circumstellar disc. The PAH carriers, however, do not reside in the
disc of these objects, but in a more recent C-rich bipolar outflow. In
\citet{gielen09} we describe some post-AGB sources where the
less frequent class C PAHs are seen. Class A PAH features are seen in
other post-AGB sources, both post-AGB stars with evidence for a
carbon-rich chemistry, e.g. IRAS\,16594-4656
\citep{garciahernandez06a,vandesteene08} as in mixed-chemistry
sources, e.g. IRAS\,16279-4757 \citep{molster99,matsuura04}. In these
two examples there is evidence for a dusty disc/torus and bipolar
outflow in which the PAH carriers reside.

If the PAH carriers reside in the disc then there has to be a
formation process where carbon-rich species can be formed in an
oxygen-rich environment. Such a scenario was proposed by
\citet{jura06} to explain PAH features observed in the oxygen-rich
giant HD\,233517. The spectrum of MACHO\,81.9728.14 does not show any
other features due to C-rich or O-rich dust species, such as SiC or
silicates, although it cannot be ruled out that these features are
present at longer wavelengths, for which no Spitzer-IRS spectrum is available unfortunately. Our chemical analysis of the photosphere shows that although
the photosphere is affected by depletion, there is no evidence for
carbon enhancement by a previous 3rd dredge-up episode.

\begin{figure}
\centering
\resizebox{8cm}{!}{\includegraphics{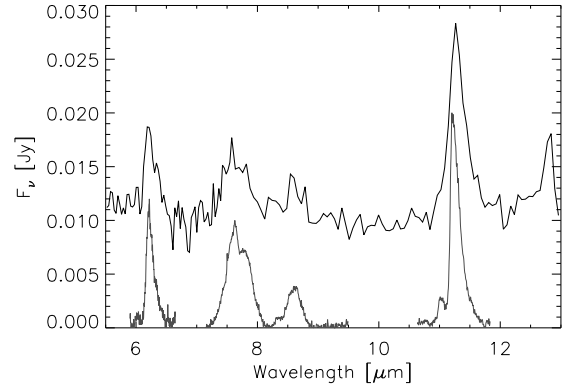}}
\caption{Continuum-subtracted spectrum of MACHO\,81.9728.14 (black solid line). Below we plot the
class A PAH emission features as described in \citet{peeters02}.}
\label{plot_pahs}
\end{figure}

\subsection{MACHO\,79.5501.13 and MACHO\,82.8405.15}

\begin{figure*}
\centering
\resizebox{4.2cm}{!}{\includegraphics{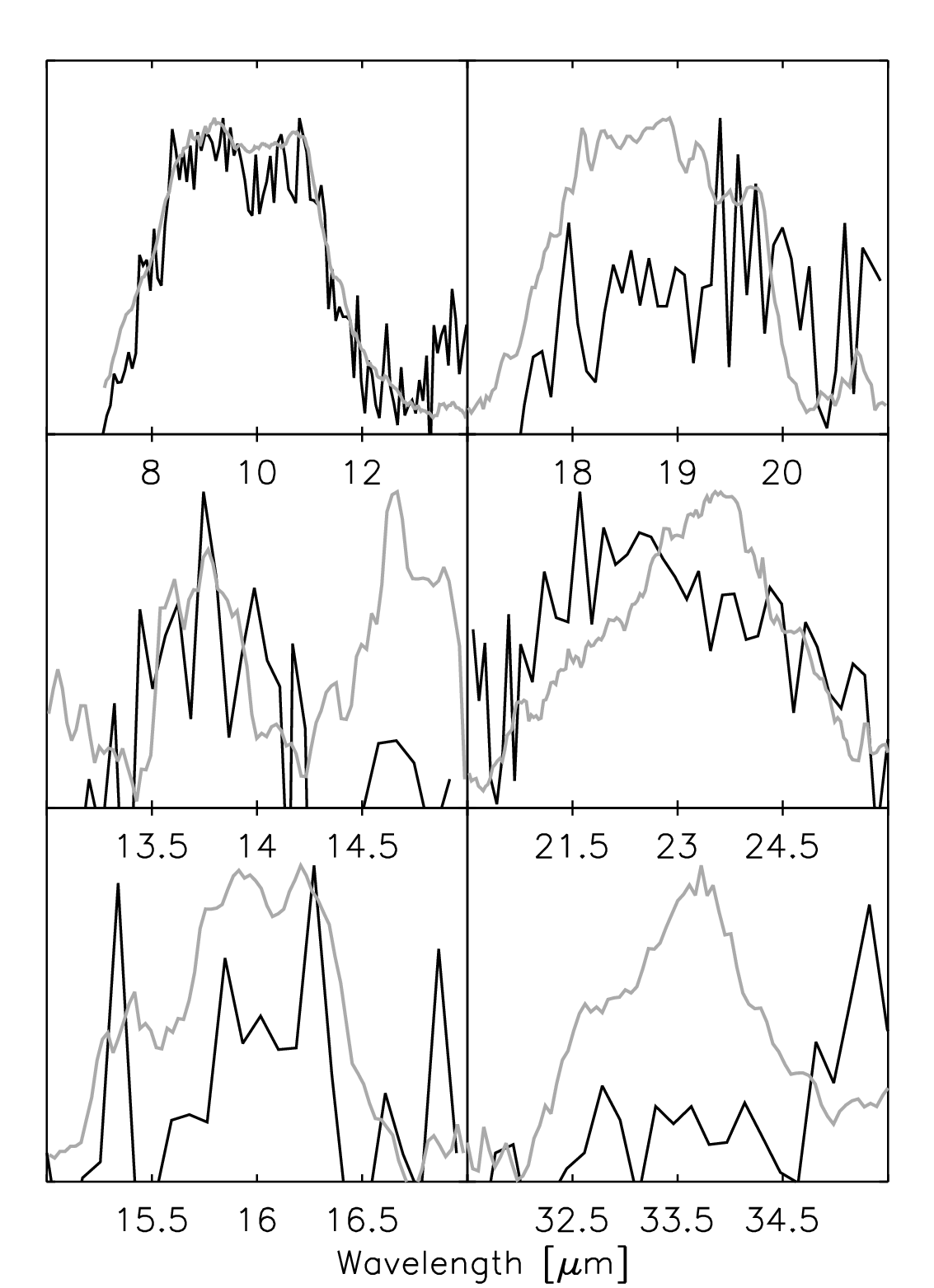}}
\resizebox{4.2cm}{!}{\includegraphics{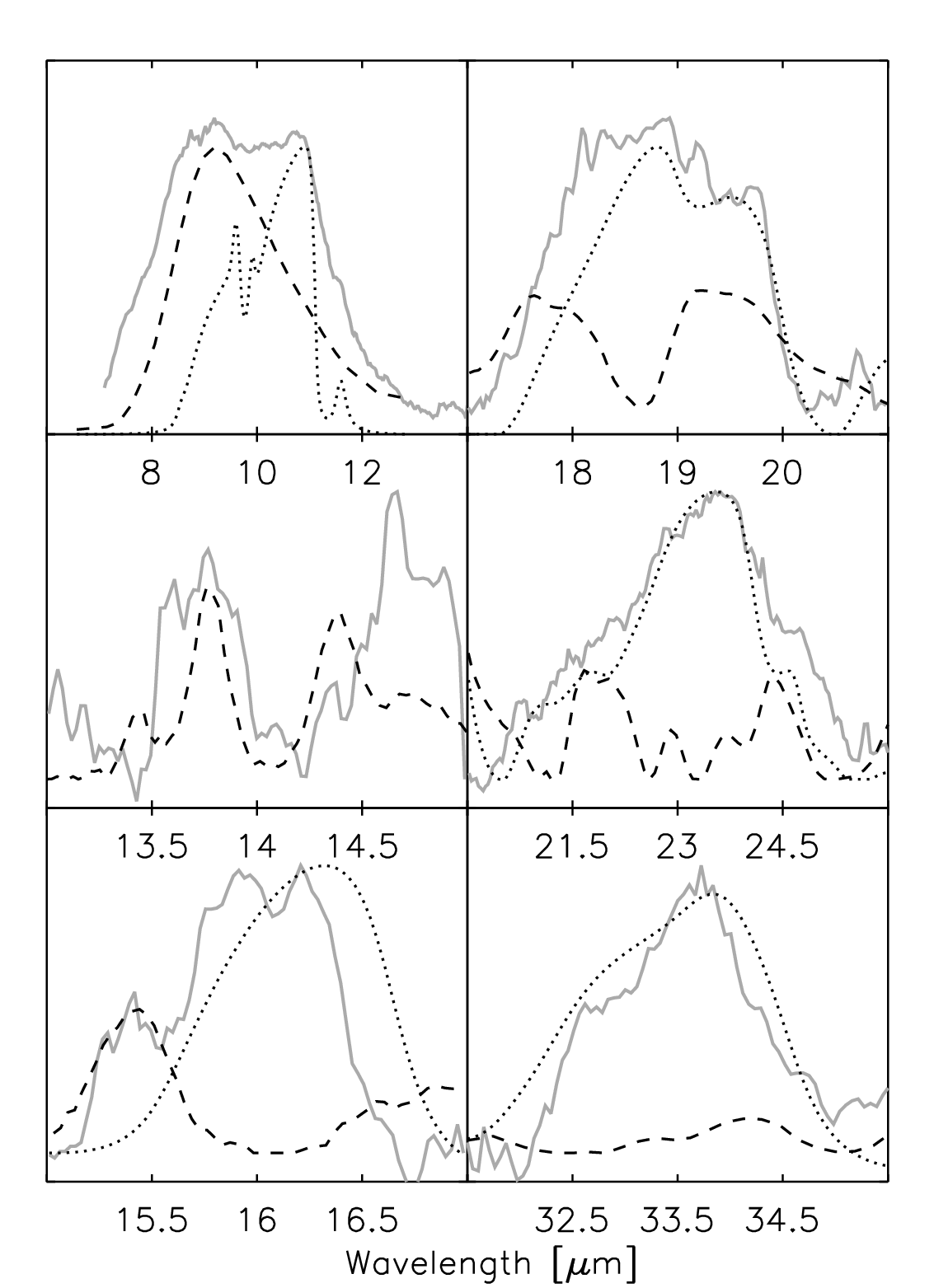}}
\resizebox{4.2cm}{!}{\includegraphics{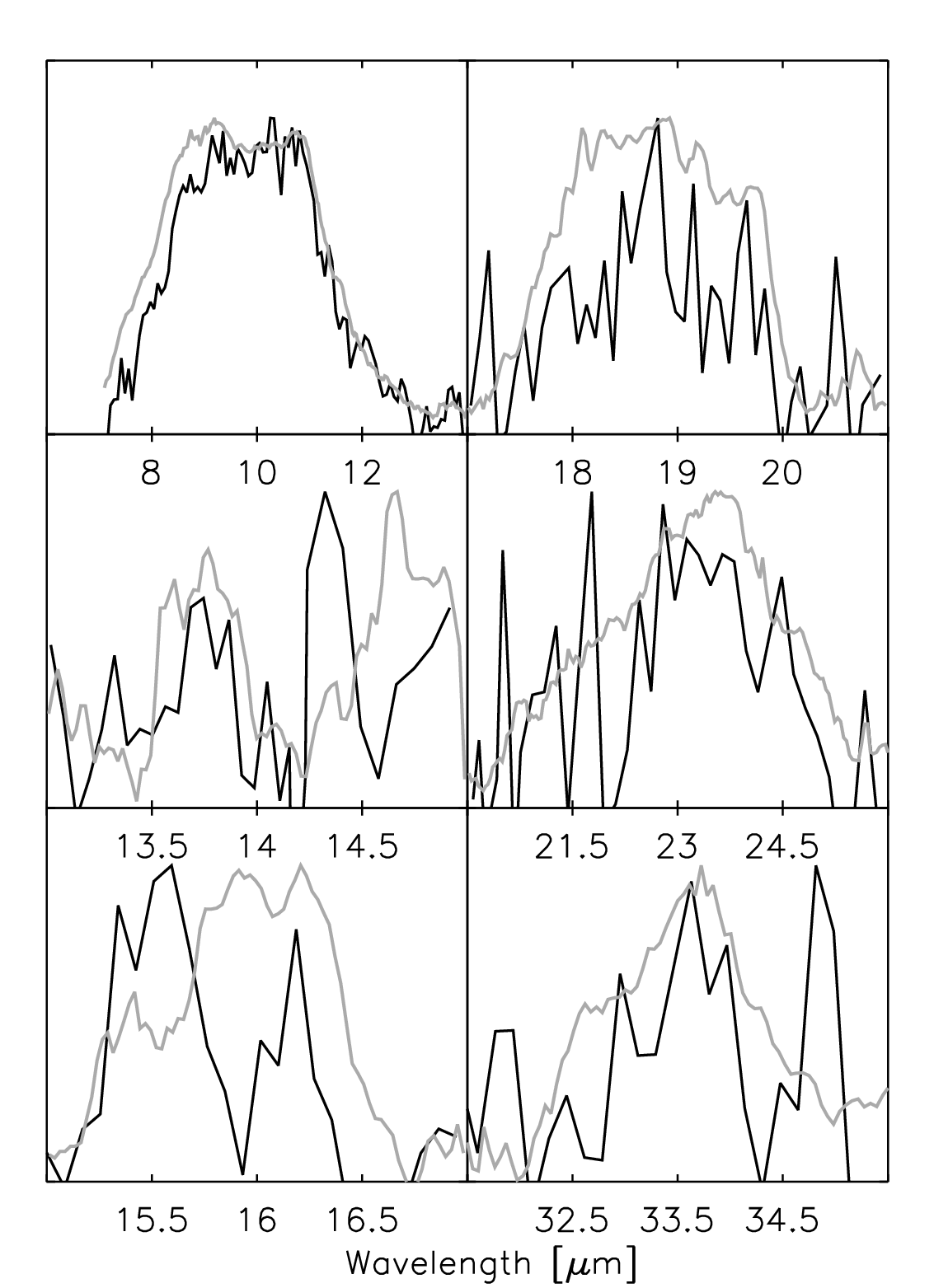}}
\caption{The $10-14-16-19-23-33$\,$\mu$m complexes of MACHO\,79.5501.13 (left), the Galactic mean spectrum (middle) and MACHO\,82.8405.15 (right), 
continuum-subtracted and normalised. Overplotted in gray is the mean spectrum of the Galactic sources \citep{gielen08}. 
For the mean spectrum of the Galactic sources we also plot the comparison
with mass absorption coefficients of forsterite (dotted), enstatite (dashed). In the 10\,$\mu$m complex
we plot amorphous olivine instead of enstatite, in dashed lines.}
\label{mean}
\end{figure*}

MACHO\,79.5501.13 and MACHO\,82.8405.15 both show clear emission
around 10\,$\mu$m. This emission feature is mainly due to amorphous
silicates, namely olivine (Mg$_{2x}$Fe$_{2(1-x)}$SiO$_4$) and pyroxene
(Mg$_{x}$Fe$_{1-x}$SiO$_3$), which peak near 9.8\,$\mu$m and
18\,$\mu$m. The observed shoulder near 11.3\,$\mu$m is likely caused by
crystalline silicate dust, namely forsterite (Mg$_2$SiO$_4$), the
Mg-rich end member of the olivine family. We cannot exclude the
presence of other amorphous silicate species, but will focus here only
on the olivine and pyroxene members.

MACHO\,79.5501.13 shows a strong slope in the continuum, and no strong
features of amorphous or crystalline silicates at wavelengths longer
than 10\,$\mu$m are detected.  Some emission might be seen around
$15-19$ and 23\,$\mu$m.  This shows that there is little cool silicate
dust in the circumstellar environment. The peak at 36\,$\mu$m is
probably a reduction residual at the end of the spectrum.

In MACHO\,82.8405.15 some small features around $15-19-23$ and
33\,$\mu$m are seen. These features can also be attributed to the
crystalline silicates forsterite and enstatite (MgSiO$_3$). The
18\,$\mu$m amorphous silicate bump is detected, albeit not very
strongly.

\subsection{Comparison with infrared spectra of the Galactic sample}

Both MACHO\,79.5501.13 and MACHO\,82.8405.15 show a strong similarity to our sample of Galactic post-AGB sources,
with similar spectral shapes and emission of amorphous and crystalline silicates.
On average the Galactic sources do show a slightly higher fraction of crystallinisation. 

In our previous study of the Galactic sample stars \citep{gielen08} we defined 6 different complexes
($10-14-16-19-23-33$\,$\mu$m complexes), and compared these to a
calculated mean complex for the 21 Galactic sources. We will use this
calculated mean Galactic spectrum to compare the observed features of
these extragalactic sources. For this we subtracted a linear
continuum from the different complexes and then normalised them. The
comparison can be seen in Figure~\ref{mean}.

The low signal-to-noise for these stars makes it hard to make a
detailed comparison, but we can already see that several features
appear in both Galactic and extragalactic sources. For both stars the
10\,$\mu$m complex is very similar to the Galactic mean complex,
although it seems slightly narrower in MACHO\,82.8405.15. The
double-peaked structure due to the presence of both amorphous
silicates and forsterite is apparent in both stars.

Around 14\,$\mu$m we do find the enstatite 13.8\,$\mu$m, and possibly
the 14.4\,$\mu$m, feature, but not the strong unidentified feature
near 14.8\,$\mu$m which is seen in the Galactic stars.

For the Galactic sample we found the forsterite 16\,$\mu$m to be
shifted to shorter wavelengths. In MACHO\,79.5501.13 we do not find
strong evidence for this feature. The peak at 16\,$\mu$m could be due
to forsterite but note that the noise is large and enhanced by the
continuum subtraction and normalisation. A strong feature at
15.5\,$\mu$m is seen in MACHO\,82.8405.15. If this is due to
forsterite the feature is shifted to even shorter wavelengths in this
star. In both stars there appear features around 19 and 23\,$\mu$m,
but only MACHO\,82.8405.15 shows a strong resemblance to the Galactic
mean feature. The 23\,$\mu$m complex in MACHO\,79.5501.13 seems to be
tilted to shorter wavelengths, instead of longer wavelengths. Whether
this points to the presence of a large enstatite fraction is unclear,
since the other enstatite features seem to be only minimal.

The 33.6\,$\mu$m forsterite feature is clearly present in
MACHO\,82.8405.15, albeit rather noisy, but can not be detected in
MACHO\,79.5501.13.

\section{Full spectral fitting}

\begin{figure}
\centering
\resizebox{9cm}{!}{\includegraphics{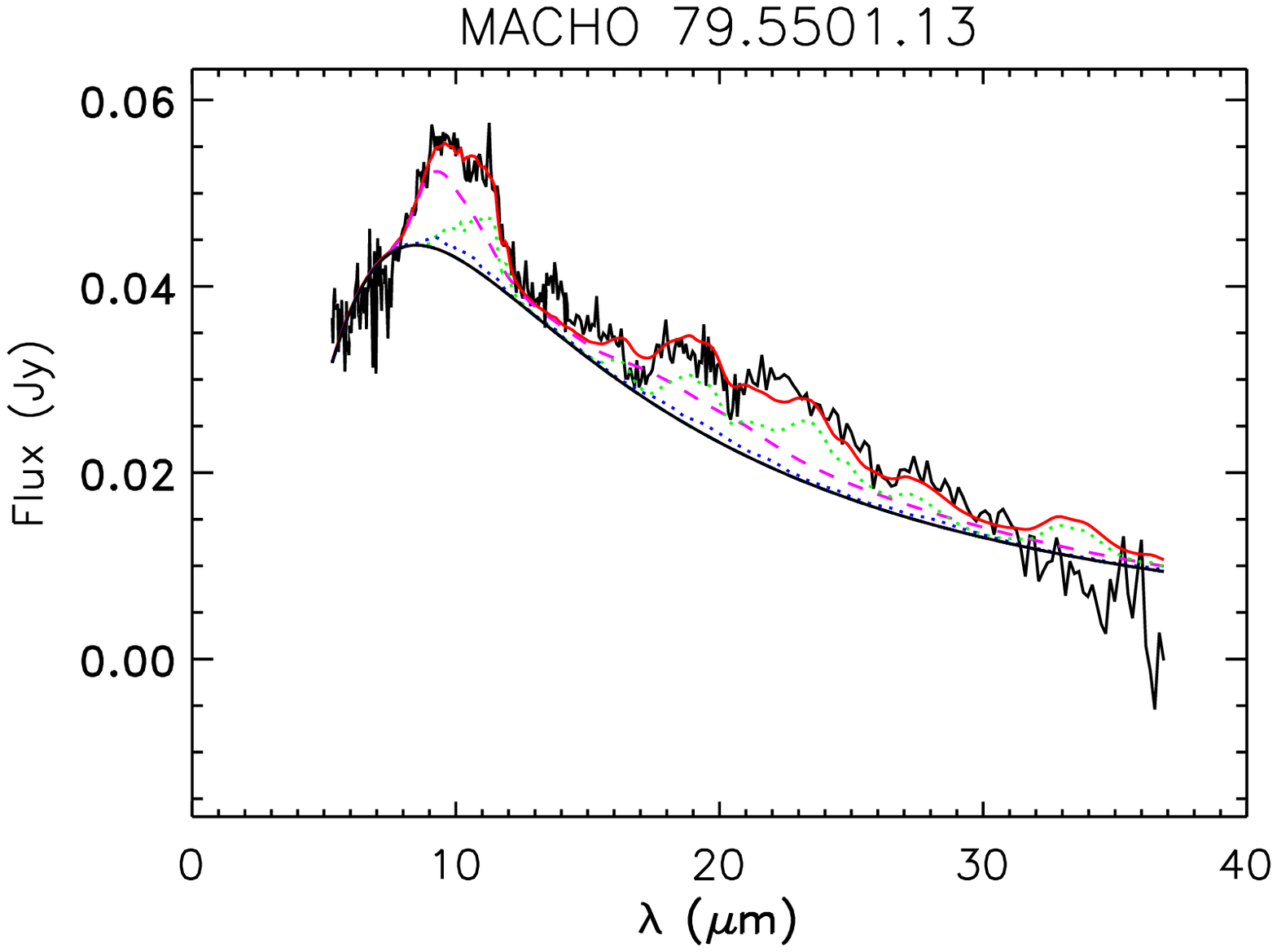}}
\resizebox{9cm}{!}{\includegraphics{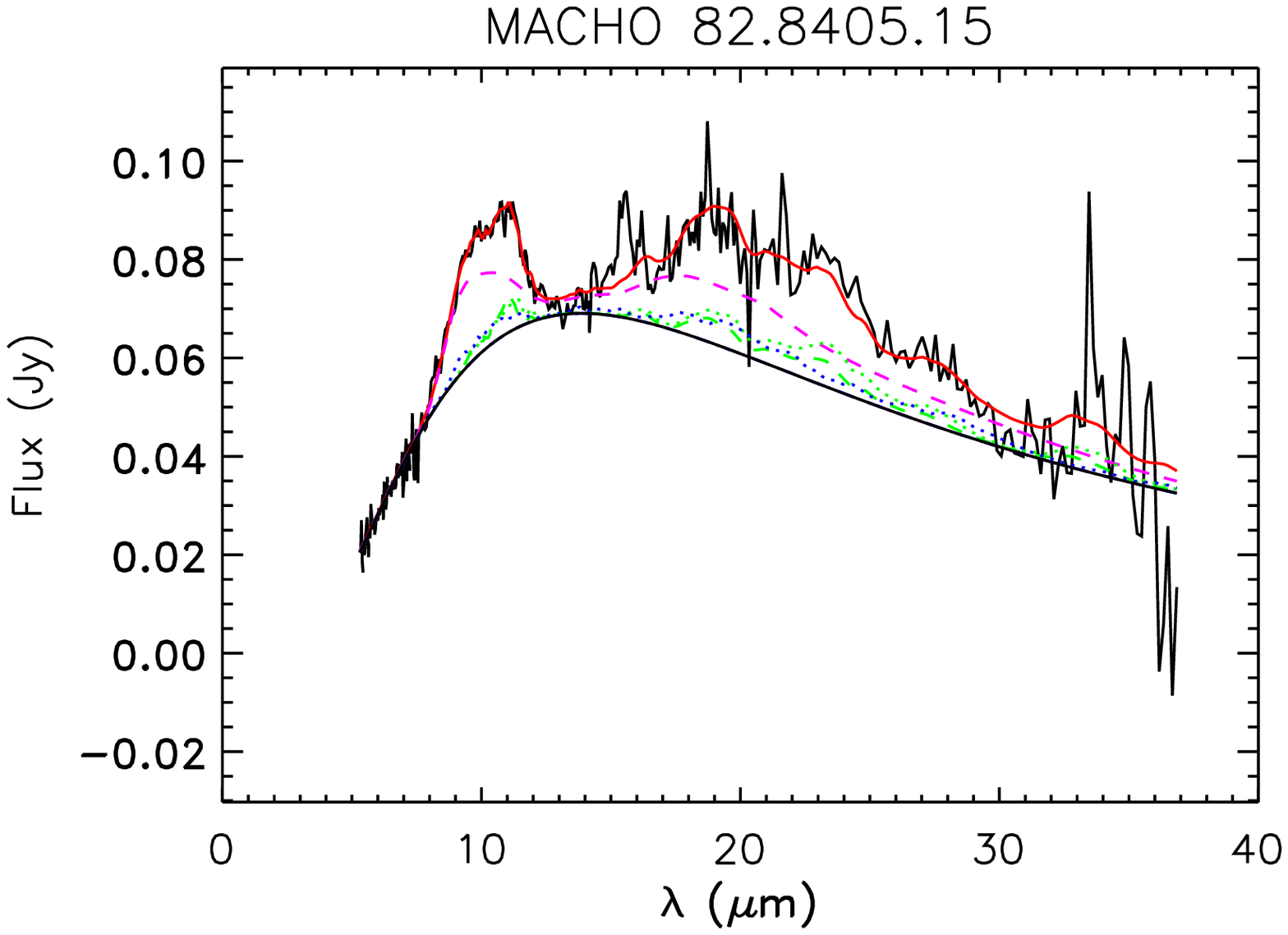}}
\caption{Best fits for MACHO\,79.5501.13 and MACHO\,82.8405.15.
The observed spectrum (black curve) is plotted together with the best model fit (red curve) and the continuum (black solid line).
Forsterite is plotted in dash-dot lines (green) and enstatite in dash-dot-dotted lines (blue).
Small amorphous grains (0.1\,$\mu$m) are plotted as dotted lines (magenta) and large amorphous grains (2.0\,$\mu$m) as dashed lines (magenta).}
\label{fullplot}
\end{figure}

For MACHO\,79.5501.13 and MACHO\,82.8405.15 we perform a full spectral
fitting as already discussed in \citet{gielen08}.  In short we assume the flux
to originate from an optically thin region, so we can make linear combinations
of the absorption profiles to calculate the model spectrum. We allow for
different dust species, grain sizes (from 0.1 to 4.0\,$\mu$m) and dust
approximations, such as Mie theory \citep{toon81}, Gaussian Random Fields \citep[GRF,][]{shkuratov05} and Distribution of 
Hollow Spheres \citep[DHS,][]{min05a} approximations. We allow for two
temperatures to describe the spectrum. Unfortunately, the lack of strong emission
features and the low signal-to-noise makes it hard to distinguish
between different models, and the calculated $\chi^2$ values are thus
quite similar. Similar to what was seen in the Galactic sources
\citep{gielen08}, we find that the best fit is obtained using
irregular grain compositions, and not a spherical description, such as
applied in
Mie theory. In Figure~\ref{fullplot} and
Table~\ref{fitresults}, we give the result of our calculated best
fit, consisting of small grains ($0.1-2.0$\,$\mu$m) in GRF dust
approximation.

For MACHO\,79.5501.13 we find a good fit to the observed spectrum.
The model continuum does not follow the strong downward trend, meaning
that the modelled continuum temperature is too high. The 10\,$\mu$m
feature is nicely reproduced, as are the small bumps near 15, 19 and
23\,$\mu$m.  The small feature around 13.5\,$\mu$m is not explained by
the model but could be due to enstatite, and was also seen in some
Galactic sources. We find the model produces a significant
33.6\,$\mu$m forsterite feature, which is not observed and thus shows
that the amount of cool dust is overestimated.

A good fit is also obtained for MACHO\,82.8405.15. Again the 10\,$\mu$m
feature and features near 20\,$\mu$m are very well fitted.  There
appears to be a rather strong feature at 15.5\,$\mu$m which is not reproduced
in the model. Forsterite has an emission feature at 16\,$\mu$m and for
the Galactic sources we already found that this features seemed to be
shifted to shorter wavelengths \citep{gielen08}, but the strength
was generally well reproduced. The feature observed here however seems
to be too strong to be explained by this shifted forsterite feature
alone.

\begin{table*}
\caption{ Best fit parameters deduced from our full spectral fitting.
}
\label{fitresults}
\centering
\vspace{0.5cm}
\hspace{0.cm}
\begin{tabular}{lrllllll}
\hline \hline
Name & $\chi^2$ &  $T_{dust1}$ & $T_{dust2}$ & Fraction & $T_{cont1}$ & $T_{cont2}$ & Fraction \\
    &          &     (K)     & (K)         & $T_{dust1}$- $T_{dust2}$    & (K)         & (K)         & $T_{cont1}$-$T_{cont2}$   \\
\hline
MACHO\,79.5501.13 & 5.1 &$ 180_{ 81}^{72}$ &$ 843_{444 }^{145}$ &$ 0.75_{0.75}^{ 0.22}- 0.25_{0.22}^{0.75 }$ &$ 458_{ 284}^{119  }$ &$ 743_{ 122}^{220  }$ &$ 0.62_{ 0.45}^{0.28}- 0.38_{0.28}^{ 0.45}$\\
MACHO\,82.8405.15  & 3.9 &$ 490_{91  }^{106  }$ &$ 737_{339  }^{207 }$ &$ 0.78_{0.69 }^{0.21 }- 0.22_{0.21 }^{0.69 }$ &$ 279_{145}^{ 59 }$ &$ 563_{ 105}^{258  }$ &$ 0.86_{0.24 }^{0.1 }-0.14_{ 0.10}^{0.24 }$\\
\end{tabular}
\centering
\vspace{0.5cm}
\hspace{0.cm}
\begin{tabular}{lccccc}
\hline \hline   
Name & Olivine & Pyroxene & Forsterite & Enstatite & Continuum\\
      & Small  -  Large & Small  -  Large &  Small  -   Large & Small  -   Large &\\
\hline
MACHO\,79.5501.13    &$3_{ 3}^{37 }    -   4_{ 4}^{48}$    &$28_{27}^{33}    -  12_{12}^{49}$    &$8_{8}^{43}    -   29_{ 25}^{24}$    &$ 8_{8}^{38} -  8_{8}^{35}$    &$87_{10}^{8}$\\

MACHO\,82.8405.15    &$ 9_{ 9}^{34 }    -   6_{6}^{60}$    &$30_{27}^{32}    -  15_{14}^{48}$    &$9_{8}^{22}    -   16_{14}^{31}$    &$7_{7 }^{32}    -  9_{8}^{28}$    &$84_{11 }^{10}$\\

\hline
\end{tabular}
\begin{footnotesize}
\begin{flushleft}
Note: Listed are the $\chi^2$, dust and continuum temperatures and their
relative fractions. 
The abundances of small (0.1\,$\mu$m) and large (2.0\,$\mu$m) grains of the various
dust species are given as fractions of the total mass, excluding the dust responsible for the continuum emission.
The last column gives the continuum flux contribution, listed as a percentage of the total integrated flux over the 
full wavelength range. The errors were obtained using a monte-carlo
simulation based on 100 equivalent spectra. Details of the modelling
method are explained in \citet{gielen08}.
\end{flushleft}
\end{footnotesize}
\end{table*}  

\section{2D disc modelling}
\label{sedmodel}

To model the energetics of the observed RV\,Tauri stars, we use the disc model as described in \citet{gielen07}, namely the
2D radiative transfer passive disc model given by
\citet{dullemond01} and \citet{dullemond04}. 
The first part of this model consists of a 2D radiative transfer code which calculates the
dust opacity and temperature. The gas component of the disc is assumed to be thermally coupled
to the dust. The second part of the model is a 1D hydrostatic equilibrium code which calculates iteratively
the vertical density profile.
Stellar input parameters are the
mass, luminosity and effective temperature of the star. For the total mass
of the system we use a value of 1\,M$_\odot$. We use the estimated
luminosity as given in Table~\ref{stelpar}, which corresponds to
a distance $d=50$\,kpc. The disc parameters consist of the disc size,
total mass, the different dust components and grain sizes and the
surface-density distribution $\Sigma(r)=\Sigma_0(r/r_0)^\alpha$, with $\alpha$ the power-law index
and $\Sigma_0$ a normalisation factor at reference distance $r_0$. The disc is made up from an
astronomical silicate dust mixture with a gas-to-dust ratio of 100.
We use a value $-2.0<\alpha<-1$ for the surface-density distribution,
as expected in a disc environment. The inclination is kept fixed at
$45^{\circ}$.

The observed feature-to-continuum ratio can be used to determine the
grain sizes in the disc, as small grains will produce much stronger
feature-to-continuum ratios than larger grains. For the Galactic
sources we already modelled \citep{gielen07,gielen09}, we opted to include an extra opacity source such as
metallic iron and/or larger grains (up to 20\,$\mu$m), but for these
extragalactic stars we need to use considerably smaller grain sizes,
ranging from 0.1\,$\mu$m to $5-10$\,$\mu$m. In the Galactic sources, we found grain growth to be very
efficient in a disc environment. This will produce dust particles with larger grain sizes, \
which will settle towards the midplane.  This
will cause the disc to be inhomogeneous, consisting of a disc made up
from small grains and a cool midplane of larger grains.

For a few stars in the Galactic sample we possessed submillimetre data
showing a blackbody slope from 60\,$\mu$m to 850\,$\mu$m, pointing to
the presence of extremely large grains, up to centimetre size or
larger. The presence of such large grains in the disc can not be
confirmed for these LMC stars since we lack data longwards of 40\,$\mu$m.

The models are calculated using the canonical gas-to-dust ratio
of 100. Since this value for the LMC may very well be higher
\citep{gordon03} because of the overall lower metallicity,
we tested how a higher ratio influences the disc structure. Increasing the
gas-to-dust ratio means that dust formation and grain growth were less
effective, and will thus give less pronounced dust signatures. We
find that using a gas-to-dust ratio of 400, we need to increase the
total disc mass by about a factor $3-4$ to get a similar dust mass,
and thus to get a similar fit to observed infrared dust
features. Changing the gas-to-dust ratio does not strongly influence
the disc structure itself so the scaleheight of the disc is not very
dependent on the gas-to-dust ratio.

As noted in \citet{gielen07,gielen09} the models are
quite degenerate and equally fitting models with slightly different
sizes, total masses and surface-density distributions can be
found. Interferometric measurements, combined with the disc model, are
invaluable to constrain further the disc geometry
\citep{gielen09}. Unfortunately, the stars are too distant and
too faint for current interferometric facilities, like the MIDI and
AMBER instruments on the VLTI.

\begin{table}
\caption{Results of our SED disc modelling.}
\label{sedresults}
\hspace{-0.3cm}
\begin{tabular}{lcccc}
\hline \hline
          & grain size & $R_{\rm in} - R_{\rm out}$ & $m$                  & $\alpha$  \\
          &  $\mu$m    & AU                         & $10^{-4}$M$_{\odot}$ &                     \\
\hline 
MACHO\,79.5501.13 & $0.1-10$ & $10-30$  & 1 & -1.5 \\
MACHO\,81.9728.14: S & $5$ & $7-100$ & 2 & -1.5 \\
MACHO\,81.9728.14: C& $0.1-10$ & $10-100$ & 0.6 & -1.0 \\
MACHO\,82.8405.15 & $0.1-10$ & $17-200$ & 6 & -1.3 \\
\hline
\end{tabular}
\begin{footnotesize}
\begin{flushleft}
Note: Given are the used grain-size distribution, the inner and outer radius ($R_{\rm in}$-$R_{\rm out}$),
the total disc mass $m$ for the homogeneous disc model and the surface-density distribution power law $\alpha$.
For MACHO\,81.9728.14 we give both the results of the silicate disc model (S), and the carbon disc model (C).
\end{flushleft}
\end{footnotesize}
\end{table}  

\subsection{MACHO\,79.5501.13}

For MACHO\,79.5501.13 we find we need a very small outer radius ($<
100$\,AU) to fit the observed SED. This could already have been
deduced from the spectrum, where the strong downward slope of the
continuum pointed to the presence of little cool dust in the disc. A
good fit was obtained by using an outer radius of 30\,AU and an inner
radius of 10\,AU. With these radii the inner rim has a temperature of
1200\,K, close to the canonical dust sublimation temperature for
silicates ($\sim 1200-1500$\,K). The result of our modelling is shown in Figure~\ref{depletie}.

We find that at longer wavelengths the observed Spitzer spectrum falls more quickly 
than the model (even within the error bars), which still overestimates the amount of
cool dust. The observed infrared spectrum at the longest wavelengths is typically less
reliable so we would need longer wavelength photometric observations to
confirm the very steep slope at longer wavelengths detected in the infrared spectrum.


\subsection{MACHO\,81.9728.14}

Modelling the SED of MACHO\,81.9728.14 using the 2D disc model is less
straightforward.  The main problem lies in the fact that we only
possess the PAH dominated Spitzer spectrum from $5-15$\,$\mu$m, and no
information at longer wavelengths. In other post-AGB sources the
observed PAH carriers do not reside in the disc but in a more recent
outflow, and so we cannot use the disc model to reproduce this part of
the spectrum. Moreover, PAHs are small and subject to single photon heating
effects, so they are not in equilibrium.

If we do try to fit the SED with a standard silicate disc, we find
that the model gives a strong 10\,$\mu$m silicate feature, which
should have been detected in the Spitzer spectrum. If the PAH features
do come from the disc, then the silicate grains must be either very large, 
to avoid producing a strong silicate feature, or very optically thick. Some small grains, however,
have to be present to explain the observed near-IR excess. Another
explanation would be that the disc is made up from another, probably
carbon-rich, dust species. The chemical composition of the photosphere
does not show evidence for a possible strong carbon enhancement.

A disc consisting of 5\,$\mu$m grains, and with inner and outer radii of
7 and 100\,AU, proves a good fit to the observed spectrum. A similar fit
however can be found using a carbon-rich disc, with grain sizes of
$0.1-10$\,$\mu$m and a disc size of $10-100$\,AU. The result of our modelling is shown in Figure~\ref{depletie}.


\subsection{MACHO\,82.8405.15}

The 2D disc model also provides a good fit to the observed SED of
MACHO\,82.8405.15. For this star we find an inner radius around
17\,AU.  At this distance the inner rim reaches a temperature of
1000\,K, slightly below the dust sublimation temperature.  The outer
radius is less well constrained, but we find that it cannot be as
small as in MACHO\,79.5501.13.  In this star we see a clear 20\,$\mu$m
silicate feature, which can only be reproduced using an outer radius
$> 150$\,AU, with a value of $\sim -1.5$ for the surface-density
distribution. The result of our modelling is shown in Figure~\ref{depletie}.


\subsection{MACHO\,81.8520.15}

The light curve of MACHO\,81.8520.15 was studied by
\citet{alcock98}. The classification of the light curve is uncertain but the
period of 42.1 days makes it, together with the calculated stellar
parameters, a possible RV\,Tauri candidate. This star has a very low
amplitude light curve with slight variability in the depth of its
minima, but the period seems to be relatively stable. \citet{alcock98}
suggest that it also might be classified as a long-period,
low-amplitude W\,Vir pulsator.

Looking at the spectral energy distribution of MACHO\,81.8520.15
(Fig.~\ref{depletie}), we see that the star has a very
small infrared excess, starting near 8\,$\mu$m. This is very different
from the typical strong infrared excesses seen in RV\,Tauri stars.

The small infrared excess shows that some circumstellar material is present,
but it remains unclear what the exact location, geometry and origin of this material is. 

Interestingly the stellar
photosphere is depleted. Photospheric chemical depletion is
the result of a process which seems to occur only in objects surrounded
by a dusty disc (see introduction), so we have at least indirect
evidence that such a stable disc must be, or must have been,
present around this object. 

The actual SED mimics that of a debris disc around younger stellar
objects. These stars are surrounded by a very evolved circumstellar
disc, which still contains trace amounts of dust and likely
planetesimals of other macroscopic objects. These disc have been cleared
of most of their original gas content. Debris discs have been found
around main-sequence stars of all spectral types \citep[e.g.][]{bryden06,chen06}.

As the photosphere is depleted, one could speculate the dust excess
seen in MACHO\,81.8520.15 as also coming from a very evolved and processed disc.
Unfortunately, the lack of more photometric or spectroscopic data at long
wavelengths makes it impossible to model the circumstellar environment of this star.

\section{Discussion}

The stellar parameters derived from high-resolution optical spectra,
combined with ample photometric data, allow for a good estimate of the
total reddening of the object. The high luminosities and the effective temperature prove the suspected post-AGB
nature of these stars.  All sample stars have effective temperatures
ranging from 5750 to 6250\,K and luminosities between 4000 and
5000\,L$_{\odot}$, assuming a typical LMC distance of 50\,kpc.  

Our chemical study shows that, also in the LMC, the photospheres of RV\,Tauri stars are
commonly affected by the depletion process. The analysis
of the Spitzer observations shows that, in the LMC as well as in the Galaxy, this depletion
process is very closely related to
the presence of a stable, dusty circumstellar disc in the system in which dust
processing has been very active.
The postulated long life-time of the dusty disc near the luminous star is a
favourable circumstance for the gas-dust separation to occur, and for
subsequent gas accretion \citep{waters92}. 

For three sample stars we possess Spitzer low-resolution infrared
spectra. Two sample stars have spectra strongly resembling our
previously studied Galactic post-AGB disc sources, with clear emission
features due to amorphous and, in lesser degree, crystalline
silicates. The atmospheric abundances, combined with the infrared spectral
characteristics and the evidence for dust processing, show that
RV\,Tauri stars in the LMC have, on average, very similar observational
characteristics to the dusty RV\,Tauri stars in our own Galaxy.

Based on our very limited sample here, the
depletion patterns as observed in the LMC differ significantly from
star to star, 
as was also observed in the Galactic sample. 

The most remarkable object in the infrared is MACHO\,81.9728.14, which is
dominated by emission peaks due to PAHs and does not show evidence for the
presence of oxygen-rich dust species. In the photosphere, there is no
evidence for an enhanced carbon abundance. The object shows some
evidence of depletion, but only marginally so. The abundance pattern is
flat with abundances around $-1.0$ and $-1.2$ and with
depletion affecting only the elements of the highest condensation
temperature. With a [C/H]\,$=$\,$-1.0$ (based on two lines), the carbon
follows this trend, and we interpret this total distribution as coming
from an object with low initial metallicity. Despite this low initial
metallicity, the dust excess is significant and the $L_{\rm IR}$/$L_{*}$
of 53\% shows that the inner rim of the dusty disc covers a wide solid
angle as seen from the star. The PAH carriers can be
categorised as being of class A, which are the strongly processed
interstellar PAHs. So far, it remains unclear whether the PAH
carriers reside in the disc, or more likely, in an outflow from the
central star. 

Also the abundance pattern of MACHO\,81.8520.15 is remarkable.  The
star shows a very high [C/Zn] ratio and a very clear correlation of
the underabundances with the condensation temperature. If we interpret
the high [C/Zn] value as due to depletion, the Zn abundance does not
reflect the initial condition, but must also be affected by gas-dust
separation and subsequent accretion.  This is remarkable as Zn is an
element with one of the lower dust condensation temperatures.  The
alternative interpretation, in which MACHO\,81.8520.15 is an
intrinsically low metallicity object and carbon is enhanced by the 3rd
dredge-up, is problematic: in such a scenario the s-process elements
should be strongly enhanced as well, which is clearly not observed.

Interestingly MACHO\,81.8520.15 has a very
small infrared excess, which seems to contradict the presence of a stable dusty
disc. However, as the star is depleted we postulate that a disc must
have been present at some time during its evolution. 
The SED does show a resemblance to the debris discs seen around young stellar objects. 
In these debris discs, grain growth has already formed
dust particles of considerable size (rocks and possibly even
planetesimals), and the gas component has been removed. If there is a disc around MACHO\,81.8520.15
which is now strongly evolved, this
would be the first time this late stage of disc evolution has been
seen around a post-AGB star. The low Zn abundance would then indicate
that the active gas-dust separation and accretion of the cleaned gas also
takes place in colder regions of the evolving disc. 


MACHO\,79.5501.13 and MACHO\,82.8405.15 are very similar in photospheric
chemical composition, SED, and in infrared spectral
appearance to the strongly depleted Galactic objects. The
underabundance in some elements is a factor of 1000 smaller
than the solar value. The initial metallicity of both objects is harder
to recover, but the flat abundance distribution of C, Zn, S and Na
suggests that both stars have an intrinsic metallicity of about
$-$0.5, which is only slightly lower than the average value of the
LMC.

The spectral energy distributions of the sample stars are well
modelled using a passive disc model. The inner radius is close to the dust 
sublimation radius for all stars, while the outer radius is poorly
constrained, but a good fit is obtained with outer radii around
$100-200$\,AU. We find, in comparison to most Galactic sources,
that there is no need for an extra opacity source, such as metallic
iron or larger grains, to reproduce the observed feature-to-continuum
ratio of the dust emission. This could be an effect of the lower metallicity of the LMC sample which reduces the
formation of metallic iron. Dust formation itself, and the subsequent
processing, has been efficient enough to produce the large scale height
needed for explaining the infrared luminosity. Unfortunately, we lack submillimetre
data to probe the component of large grains.
Individual crystalline silicate bands do show
some differences from the Galactic mean spectrum. To deduce whether this
is a general trend in the LMC, and thus points to a different dust
composition, grain size or dust model, we need to considerably enlarge
the LMC sample. This will be done in a future study. This extension of the sample
will hopefully give a wider range in evolutionary phases in the post-AGB history of these 
sources, which we then hope to relate to other observed parameters, such as element abundances,
disc parameters or infrared mineralogy.

\section{Conclusions}

Clearly the formation of stable dusty discs and the significant
feedback of this disc on the central star is
not exclusive to our Galaxy alone. Four out of five LMC RV\,Tauri
objects, for which we have high-resolution data, are found to be strongly
affected by the depletion process in which the atmospheres became poor
in refractory elements. Moreover, the infrared colours and spectral
data show that three sources are surrounded by a highly processed,
stable disc in which dust processing has been efficient. In one
source PAH particles are formed, while there is no other evidence for
intrinsic carbon enhancement in the photosphere. This object is
intrinsically the most metal poor object of the sample. For MACHO\,81.8520.15 
we only found a small dust excess at 8\,$\mu$m and we could interpret this
as evidence for strong disc evolution. In this star, even the Zn abundance
is affected by depletion which means that the gas-dust separation
occurred at low temperatures.

Overall the RV\,Tauri stars in the LMC display many characteristics of
their Galactic peers. Whether these extragalactic stars also reside in
a binary system remains unclear with the available data at hand. The low magnitude of these
stars does not allow for a long-term radial velocity monitoring
programme but, given the strong similarities to the Galactic binaries,
a binary evolutionary channel is very likely needed to understand the RV\,Tauri
stars in the LMC as well.

\begin{acknowledgements}
CG and HVW acknowledge support of the Fund for Scientific Research of Flanders
(FWO) under the grant G.0178.02. and G.0470.07. This work is based on observations
made with the Spitzer Space Telescope, which is operated by the Jet Propulsion Laboratory,
California Insitute of Technology, under a contract with NASA.
\end{acknowledgements}

\bibliographystyle{aa}
\bibliography{/STER/100/cliog/disk28/Artikels/referenties}

\end{document}